*Research Article*

# Conception of Biologic System: Basis Functional Elements and Metric Properties

**Garri Davydyan**

*Appletree Medical Group, 1902 Robertson Road, Ottawa, ON, Canada K2H 7B3*

Correspondence should be addressed to Garri Davydyan; garri.davydyan@gmail.com

A notion of biologic system or just a system implies a functional wholeness of comprising system components. Positive and negative feedback are the examples of how the idea to unite anatomical elements in the whole functional structure was successfully used in practice to explain regulatory mechanisms in biology and medicine. There are numerous examples of functional and metabolic pathways which are not regulated by feedback loops and have a structure of reciprocal relationships. Expressed in the matrix form positive feedback, negative feedback, and reciprocal links represent three basis elements of a Lie algebra $sl(2, \mathbb{R})$ of a special linear group $SL(2, \mathbb{R})$. It is proposed that the mathematical group structure can be realized through the three regulatory elements playing a role of a functional basis of biologic systems. The structure of the basis elements endows the space of biological variables with indefinite metric. Metric structure resembles Minkowski's space-time $(+, -, -)$ making the carrier spaces of biologic variables and the space of transformations inhomogeneous. It endows biologic systems with a rich functional structure, giving the regulatory elements special differentiating features to form steady autonomous subsystems reducible to one-dimensional components.

## 1. Introduction

The concept "system," introduced in the middle of the last century, was proposed to describe self-organizing properties of biologic matter. The essence of it is the ability of biologic objects to maintain their own anatomical and functional structure [1–3].

Until now positive and negative feedback are used to describe the regulatory mechanisms of biologic system. Existing data show that positive and negative feedback do not explain some of the physiological and medical data. It was an attractive idea to find out whether some regulatory components can form a functional basis of the system. In challenging the existing hypotheses it was found that reciprocal links have the same importance as positive and negative feedback in functional regulation of biological systems. In fact, positive and negative feedback and reciprocal links can form a functional basis of biological systems.

Regulatory functions of the system comprise numerous metabolic pathways and biochemical reactions, which, in fact, realize only a number of core mechanisms. Until now a great deal of effort has been undertaken to find these functional invariants. Positive and negative feedback are the only known functional structures with universal properties [4–8].

Widely presented as common regulatory mechanisms in general biology and clinical practice to explain normal physiological and pathophysiological mechanisms, feedback loops do not correspond to counter-directional (reciprocal) pathways.

There are some examples showing broad representation of reciprocal relationships in regulation of metabolism and basic biologic functions. The system of hemostasis consists of two "antagonistic" subsystems, clot formation and clot degradation cascades; concentration of the glucose in the blood is regulated by insulin and glucagon releasing mechanisms leading to the opposite results. Parathyroid hormone increases and calcitonin decreases calcium concentration in the blood. The natriuretic peptide is opposite to the renin in vasopressor function. It diminishes the blood pressure, whereas renin increases it. Morphogenesis and apoptosis also follow the reciprocity as their leading regulatory mechanisms. Less trivial examples provide subsystems, which are not opportunistic or, figuratively speaking, not lying on the line of plus-minus relationships but rather have complementary properties and are "orthogonal." These are androgen and estrogen promoted biochemical pathways leading to the different phenotypical and functional properties [5].

Biological meaning of the reciprocal structures is that each pair of functions, linked reciprocally, seems to be originated from a common morphological and functional root (or stem). Reciprocity implies the link between two opportunistic, differentiated from common functional root, functions.

Negative feedback is well known in general and reproductive endocrinology because of its quite demonstrative mechanism balancing activities of central and peripheral endocrine organs. For example, lack of peripheral hormones signals central organs to produce more stimulating substances. Stimulating factors activate peripheral systems, which increase production of peripheral hormones. If endocrine glands produce an excessive amount of peripheral hormones, it blocks central stimulation, which, in turn, diminishes peripheral activity.

Positive feedback shows its action in clot formation cascades. Another example of positive feedback is pulse therapy. Increasing the doses of hormones administered at short intervals improves functional capabilities of both peripheral and central systems. Positive feedback loops also help to explain quickly developing emotional reactions like acute phobias. In reproductive endocrinology, in the first stage of the delivery, release of the oxytocin stimulates uterus contractions, cervix dilatation, and effacement, which, in turn, stimulates oxytocin release. This vicious circle is maintained by positive feedback until the process switches to another regulatory mechanism corresponding to the second stage of delivery [5, 9].

Based on these observations, it is proposed that "reciprocal links" (reciprocity) as well as positive and negative feedback are the major determinants and separable functional elements of the internal self-regulatory structure of the system. Treated as separable functional elements, positive and negative feedback and reciprocal links expressed in the matrix form are identical to the basis of elements of Lie algebra sl(2, $\mathbb{R}$) of special linear group SL(2, $\mathbb{R}$); $\mathbb{R}$ represents real numbers.

Lie algebra sl(2, $\mathbb{R}$) and the related group SL(2, $\mathbb{R}$) are *closed sets* of elements endowing positive and negative feedback and reciprocal links with some additional functional properties. It includes the ability to form integrative units through the linear combinations of matrices representing these elements [10].

Formally from the structure of the chosen basis elements of the Lie algebra it follows that the space of biologic variables has indefinite metric, meaning that the space of possible regulatory scenarios is inhomogeneous. Indefinite metric gives a broad spectrum of regulatory actions determining system's behavior.

## 2. Graph Representation of Functional Structure of a System

According to the classical system's theory, a system is a set of interrelated elements, acting together as a whole to achieve a specific outcome for the given system. As a whole a system has its own global or external function and in this sense it acts as a machine, transforming input into output. (Further in the text a *Biologic System* is a general term applied to different morphologically and functionally separable objects of biological origin. A simplest two-element system is defined as a pair $(L, \mathcal{G})$, where $L$ is a space of biologic variables, realized as $\mathbb{R}^2$, and $\mathcal{G}$ is a space of transformations on $L$ realized as $2 \times 2$ matrices over $\mathbb{R}$. $\mathcal{G}$ can be represented either by **G**, which are $2 \times 2$ (2 times 2) matrices over $\mathbb{R}$ with determinant one, or $\mathfrak{g}$ which are linear approximations of **G**, represented by $2\times 2$ matrices of trace (the sum of diagonal elements) zero.) The input-output relationship does not provide any information about internal regulatory elements and their structure. The simplest way to express internal structure of the system is to show it as a directed cycle graph where points are internal elements and arrows are functional links among them.

Biomolecules, cells, tissues, organs, and so forth are some examples of the *elements* of biologic systems. *Relations* among the elements can be "visualized" through the chemical affinity of molecules triggering biochemical reactions, or even through the generated physical forces applied to the tissues and organs. Detailed functional structure of a system shown by a web (simple graph) of linked to each other elements usually is not observable, while generalizations may hide and lose some substantial functional properties. Common functional or regulatory characteristics inherent to different biological objects can be presented, for instance, by closed loops indicating self-regulatory functional structure.

Most of the proposed regulatory mechanisms and theories of functional organization of biologic systems are based on negative feedback loops [4], which can be demonstrated in the following example. For instance, digestive systems in protozoa and animals are utterly different, but, at the same time, they have some common features that aimed to capture sources of energy, digest, and move partially digested substances until they can be excreted. In protozoa digestion is rudimental, proceeded not in a specialized tract but in the cytosol. During digestion captured substances form electrochemical and mechanical gradients, which create forces moving the content within the cell. This process seems to be continually adjusted by feedback loops through the sensory stimuli coming from every movement of being digested particles. By feedback loops the system forms an "optimal" trajectory, which mimics temporarily created digestive tract.

Contrary to the primitive unicellular organisms, the passage of chime in animals is implemented by specialized organ-gastrointestinal tract. Despite the complex functional and anatomical organization, motility mechanism of the intestine seems to also involve negative feedback as a key regulatory component. It certainly follows from the existence of closed filling-emptying cycles as possible regimes of regulation of the motor function of intestinal segment [11].

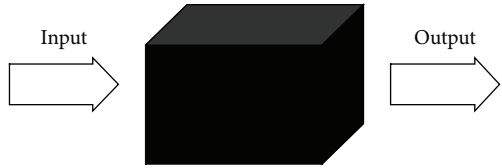

Figure 1: The "black box" represents undetermined internal structure of a system.

By now there are no conceptual models describing regulatory structure of the system realized in the *space* of functional elements [8]. We still apply to the classical image of a system, a "black box," because of undetermined system's internal functional structure (Figure 1).

The graph representation gives an initial approach in understanding of the basic structural differences among positive feedback, negative feedback, and reciprocal links. Schematically, the simplest *internal functional structure* is given by a two-point closed graph: points are two subsystems, and arrows, having plus or minus signs, are the links. Each pair of elements and arrows with fixed signs gives one functional (regulatory) unit. For example, negative feedback has arrows having plus and minus signs, while positive feedback has arrows with only positive signs. This scheme shows how each subsystem responds on exiting (+), inhibiting (−), or neutral (0) stimuli. Formally, there exists over sixteen scenarios, assuming that each subsystem is also sending the signals to itself. Consider two-element graph, when each element is directly linked only to itself and signs of the loops are opposite. If "neutral," indirect, links are assumed to be between elements, this graph will represent reciprocal (antipodal) functional structure. In the reciprocal graph elements do not affect each other directly, so the links cannot be shown by +/− arrows and are denoted as neutral (Figure 2). However, information received through the neutral pathways eventually will counterbalance functional states of the elements and the whole system. For example, functional system responsible for hemostasis consists of two parts: clot formation and clot degradation subsystems. They belong to the same hierarchical level and targeted at the only one physiological parameter—viscosity of the blood. Each subsystem contributes in an opposite way. Together they provide a wide spectrum of regulatory responses to maintain the most favorable rheology for metabolic needs.

Most probably, clot formation and clot degradation subsystems were originated from the single anatomofunctional root or stem. From the stem of inseparable characteristics the process of differentiation was followed by a long way of functional and morphological splitting. As a result, some functional components became partially independent (autonomous), comprising their own cascades of biochemical reactions aimed to the opposite goals—the clot formation and the clot degradation. On the other hand, together they provide an efficient mechanism in achieving the optimal viscosity of the blood. Logically, clot formation and clot degradation biochemical pathways should be functionally linked.

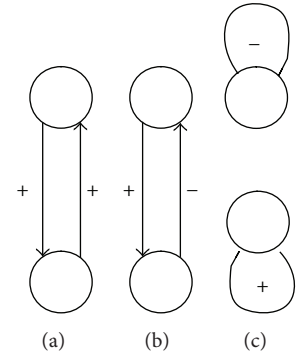

Figure 2: Graph representation of positive feedback (a), negative feedback (b), and reciprocal links (c).

Two-element graphs showing positive and negative feedback mechanisms have found practical applications mostly in endocrinology, because of the simple and convincing demonstration of two-directional relationships between central and peripheral endocrine organs. On the contrary, the graph structure of the reciprocity does not give a satisfactory picture. The logical schemes have one substantial disadvantage—they do not show regulatory structure as a dynamic process. In addition, regulatory elements should possess some integrative properties and be able to maintain functional stability of the system.

## 3. Classical Dynamic Modeling of Internal Functional Elements of a System

Ordinary differential equations (ODE) give a unique means to demonstrate dynamic properties of the objects described by variables changing in time. In ODE positive and negative feedback and reciprocal links are represented by the operators (Linear operators and matrices of linear operators are isomorphic objects with almost the same properties. A matrix is an operator expression relative to some basis. For simplicity, they will be used as synonyms, if the differences are not emphasized. All matrices, if not specially mentioned, are expressed relative to the standard basis.) (or matrices) relating velocities of variables with variables themselves. Qualitative solutions of ODE can be shown by curves, which are evolutions of the system's conditions in time.

For example, three simple differential equations

$$\begin{aligned}\frac{d\mathbf{u}}{dt} &= A\mathbf{u}, \\ \frac{d\mathbf{v}}{dt} &= B\mathbf{v}, \\ \frac{d\mathbf{w}}{dt} &= C\mathbf{w}\end{aligned} \quad (1)$$

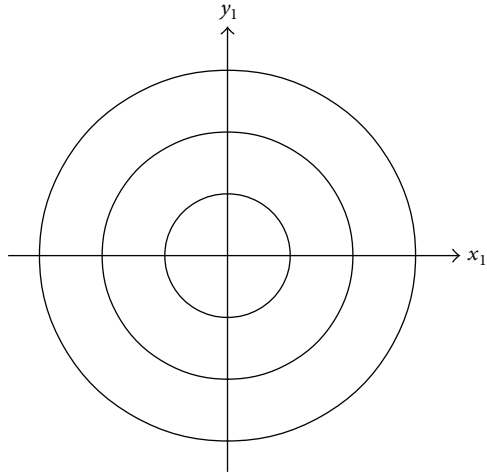

Figure 3: Phase curves of differential equation $d\mathbf{u}/dt = \mathbf{A}\mathbf{u}$. Matrix $\mathbf{A} = \begin{pmatrix} & 1 \\ -1 & \end{pmatrix}$ represents negative feedback.

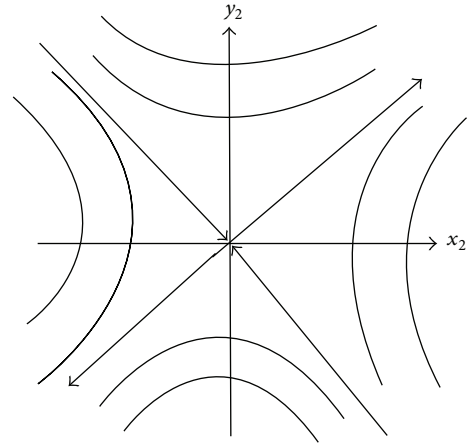

Figure 4: Phase curves of differential equation $d\mathbf{v}/dt = \mathbf{B}\mathbf{v}$. Matrix $\mathbf{B} = \begin{pmatrix} & 1 \\ 1 & \end{pmatrix}$ represents positive feedback.

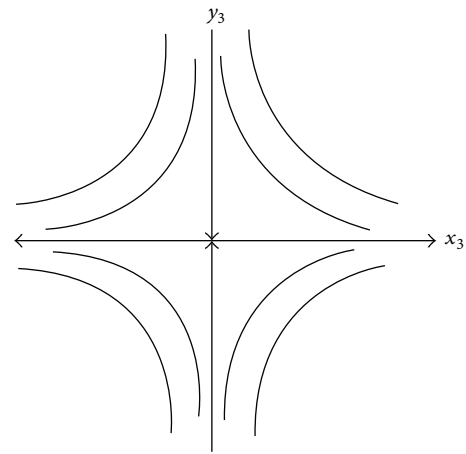

Figure 5: Phase curves of differential equation $d\mathbf{w}/dt = \mathbf{C}\mathbf{w}$. Matrix $\mathbf{C} = \begin{pmatrix} 1 & \\ & -1 \end{pmatrix}$ represents reciprocal links.

may represent positive and negative feedback and reciprocal links (Figures 3–5) in two-element systems, where the system's elements $\mathbf{u} = (x_1, y_1)$, $\mathbf{v} = (x_2, y_2)$, and $\mathbf{w} = (x_3, y_3)$ (vectors) are variables of ODE. Because two variables $(x_i)$, $(y_i)$, $i = 1, 2, 3$, are used for each system, the spaces of variables are two-dimensional, and the operators $A$, $B$, and $C$ are represented by $2 \times 2$ matrices with real coefficients $\mathbf{A} = \begin{pmatrix} & 1 \\ -1 & \end{pmatrix}$, $\mathbf{B} = \begin{pmatrix} & 1 \\ 1 & \end{pmatrix}$, and $\mathbf{C} = \begin{pmatrix} 1 & \\ & -1 \end{pmatrix}$. The matrix view directly follows from the graph representation of the functional elements. The operators $A$, $B$, and $C$ serve as the symbolic equivalents of positive and negative feedback and reciprocal relations, because the described processes are entirely determined by these operators. Each ODE has two variables $(x_i)$, $(y_i)$, $i = 1, 2, 3$; hence, functional relations between variables can be shown on the plane $(x_i, y_i)$, $i = 1, 2, 3$. Conditions of the system are points on the plane, evolving in time, along the third, the time, and axis, so that three-dimensional processes (integral curves) are shown as the projections of integral curves on the plane (phase curves). The curves characteristics are determined by the types of the matrices, so that each matrix establishes its own dynamic law of variables behavior.

Trajectories corresponding to the negative feedback (operator $A$) (Figure 3) show fluctuations of the variables, which remain within some restricted areas. Closed curves clearly demonstrate self-regulatory process [11, 12].

On the contrary, positive feedback (operator $B$) changes conditions hyperbolically (Figure 4). This behavior, if not corrected, will destroy the system. Reciprocal links (operator $C$) are also represented by hyperbolas, where the axes of variables are asymptotes (Figure 5). This type of regulations is not self-destructing. Interestingly, each point on the hyperbolas gives the same product $x_3 y_3 = \text{const}$, meaning that the increase of the value of any variable simultaneously decreases the value of another, so that the product $x_i y_i$ remains unchanged. It can be hypothesized that evolution of the system's conditions determined by operator C leaves intrinsic characteristic of the system (subsystem) unchanged. which also leaves intrinsic characteristic of the system (subsystem) unchanged, if they measured by the scalar product. For example, components of input can be entirely used in output, or there is no dissipation of the inner energy during transformations and the energy is preserved in the final product and so on.

Unlike operators $A$ and $B$, $C$ has a diagonal matrix. Related to this matrix operator $C$ has a special property; it divides 2-dimensional space of variables on two one-dimensional *invariant subspaces*, which in this case are coordinate axes.

Operator $C$ makes the subspace of each variable separable and autonomous. It acts on the vectors of invariant subspaces (eigenvectors) by changing their lengths or directions to opposite, leaving them in the same subspaces. If initial condition of the system lies on coordinate axis, it will evolve along this axis.

All linear second order ODE can be grouped on four classes of equivalence. Each class is represented by the matrix and has, related to this matrix, a phase portrait. Matrices **B** and **C** belong to the same class, because they can be transformed to each other. Their phase portraits are *the saddles* turned on 45° to each other. Matrix **A** corresponds to the phase trajectories termed *the center*. The center and the saddle cannot be transformed to each other by continuous transformations, so that **A** is not similar to **B** or **C**. There are no means **A** to be smoothly transformed to **B** or **C**. From the matrix properties it follows that negative feedback is not typically (physiologically) transformable to positive feedback or reciprocal links, suggesting some bypassing mechanisms bridging these elements. Matrices **A**, **B**, and **C** are not singular and have inverse ones. It means that describing processes can be reversed.

Properties of dynamic systems based on the described three matrices are discussed in detail in numerous publications devoted to biologic and medical problems [12, 13]. However, by now it is not understood whether regulatory processes involve all three (or only positive and negative feedback) functional components simultaneously or there is a consequence of alternating regulatory commands depending on the current conditions. Another question is how many functional units are involved in regulatory process and whether their superposition is possible [6].

## 4. Functional Structure of a Biologic System Follows Lie Group $SL(2, \mathbb{R})$ and Lie Algebra $sl(2, \mathbb{R})$ Properties

It is known that every physiologic variable fluctuates around some equilibrium states. It seems logical to assume that regulatory mechanisms preventing structural deterioration should provide reversibility of the current nonequilibrium conditions in order that the process may evolve towards the equilibrium state.

Generally, reversibility is a result of the counter regulatory mechanisms, which makes functional states fluctuate around the equilibrium points. These fluctuations are observed for the most of physiologic variables. Based on these observations, it is proposed that regulatory process reflected in fluctuations of physiological parameters possesses mathematical group properties. (A mathematical group is a set of elements of some origin satisfying the following conditions: composition of any two elements of a group belongs to the group ($r, s \in G \rightarrow r \circ s \in G$); each element has inverse one ($q \in G \rightarrow q - 1 \in G$); there exists a unique element of the group 1, which is a neutral element $q^{-1}q = 1$. Under the law of composition the group is closed structure.) A formal structure of the group requires a *set* of elements of the group and a *law of composition* of the elements.

Many functional models use positive and negative feedback as regulatory mechanisms of two component biologic systems [6, 7]. However, positive and negative feedback have never been discussed in terms of *functional basis elements* of a system.

If the basis is proposed, it infers the existence of the *space* of functional elements, composed of the basis. It also implies a rule how to compose elements of this space, including the basis. Biological meaning of group structure is that at each time point the regulatory process is determined by the functional unit formed of the basis regulatory elements.

In the standard basis $\mathbf{e}_1 = \begin{bmatrix} 1 \\ 0 \end{bmatrix}$, $\mathbf{e}_2 = \begin{bmatrix} 0 \\ 1 \end{bmatrix}$ of the carrier space of biologic variables $L$, positive and negative feedback and reciprocal links are realized as matrices corresponding to the basis elements

$$\mathbf{S}_0 = \begin{pmatrix} & 1 \\ -1 & \end{pmatrix}, \qquad \mathbf{S}_1 = \begin{pmatrix} 1 & \\ & -1 \end{pmatrix}, \qquad \mathbf{S}_2 = \begin{pmatrix} & 1 \\ 1 & \end{pmatrix} \qquad (2)$$

(Matrices of operators $A$, $B$, and $C$ are the same as $\mathbf{S}_0$, $\mathbf{S}_1$, and $\mathbf{S}_2$.) of the Lie algebra $sl(2, \mathbb{R})$, denoted $\mathfrak{g}$, of special linear group $SL(2, \mathbb{R})$.

The Lie algebra $sl(2, \mathbb{R})$ and the Lie group $SL(2, \mathbb{R})$ are closely related structures. $SL(2, \mathbb{R})$ is a multiplicative group of second order ($2 \times 2$) matrices over $\mathbb{R}$, with determinant one. It represents a group of transformations of the states of a biologic system. As a mathematical object $SL(2, \mathbb{R})$ is a smooth surface (manifold), biologic transformations of the current states of the system can be described in terms of the smooth curves rather than discrete points separating one state from another. Unit matrix is a neutral element of the group. Multiplication of any two matrices of $SL(2, \mathbb{R})$ gives element of the group. Each matrix has its inverse. Because each $2 \times 2$ matrix over $\mathbb{R}$ belongs to $\mathbb{R}^4$, determinant one restricts the four-dimensional space of transformations to three dimensions, so that $SL(2, \mathbb{R})$ is a three-dimensional manifold.

Lie algebra $\mathfrak{g}$ is a linear approximation of $SL(2, \mathbb{R})$. It is an additive subgroup of all matrices of the trace (the sum of diagonal elements) zero. The Lie algebra $sl(2, \mathbb{R})$ is also a linear vector space over $\mathbb{R}$ of the second order traceless matrices. This vector space is a tangent space $T_g$ to each point $g$ (matrix) of $SL(2, \mathbb{R})$. Tangent vectors can be moved to the identity point of the group, so that the tangent space at the identity contains all the algebra elements. Matrices $\mathbf{S}_0$, $\mathbf{S}_1$, and $\mathbf{S}_2$ lie on the tangent space at the identity to the Lie group $SL(2, \mathbb{R})$ and form a basis of three-dimensional vector space $\mathbb{R}^3$. Linear combinations of basis elements with real coefficients $\mathbf{S} = a\mathbf{S}_0 + b\mathbf{S}_1 + c\mathbf{S}_2$, ($a, b, c \in \mathbb{R}$), give elements of $\mathfrak{g}$, $\mathbf{S} \in \mathfrak{g}$, $\mathbf{S} \in \mathbb{R}^3$. The $\mathbf{S}_0$, $\mathbf{S}_1$, and $\mathbf{S}_2$ matrices are not singular and have the inverse ones.

Neighborhood of $T_g$ to each point $g$ gives a linear approximation of the surface $SL(2, \mathbb{R})$ surrounding this point. Locally, real objects are qualitatively similar to their linear approximations. If there is a curve on $SL(2, \mathbb{R})$ its behavior can be described in the neighborhood of each point of this curve by elements of the Lie algebra lying on the tangent surfaces to these points (Figure 6).

As a Lie algebra, $sl(2, \mathbb{R})$ has a special bilinear operation— not symmetrical (depending on the order of elements) Lie bracket [ , ] defined as $[\mathbf{X}, \mathbf{Y}] = \mathbf{XY} - \mathbf{YX}$, where $\mathbf{XY}$ ($\mathbf{YX}$) is a matrix multiplication. For each two elements of the algebra $\mathbf{X}$ and $\mathbf{Y}$, result of the bracket operation is also an element of the algebra $[\mathbf{X}, \mathbf{Y}] \in sl(2, \mathbb{R})$, so $sl(2, \mathbb{R})$ is closed under the Lie bracket. Bracket operation is a criterion for the matrices to belong to the algebra. It also measures symmetry (commutativity) of matrix products ($\mathbf{XY}$ versus $\mathbf{YX}$). If matrices commute, then the result of bracket operation is zero; $\mathbf{XY} = \mathbf{YX}$. sl(2, ℝ) is not a commutative group; that is, multiplication in the algebra sl(2, ℝ) is not a symmetrical operation, which means that the result depends on the order of multiplying matrices.

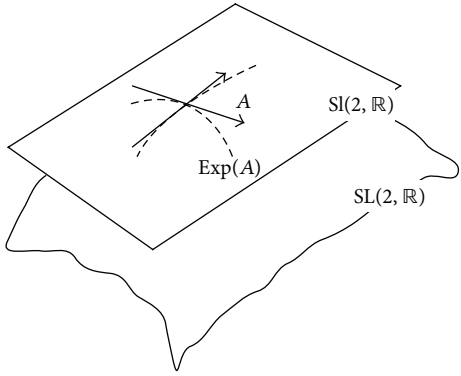

Figure 6: The space of transformations SL(2, ℝ) of biologic variables and its linear approximation tangent 3-space sl(2, ℝ).

Lie brackets of basis matrices are $[\mathbf{S}_0, \mathbf{S}_1] = -2\mathbf{S}_2$, $[\mathbf{S}_0, \mathbf{S}_2] = 2\mathbf{S}_1$, $[\mathbf{S}_1, \mathbf{S}_2] = 2\mathbf{S}_0$. In terms of transformations of functional states noncommutativity means the following: the resultant point of transformation $\mathbf{XY}$ ($\mathbf{X}$ following $\mathbf{Y}$) is not the same as for $\mathbf{YX}$. Elements of the group can be obtained by exponential mapping of the algebra elements, exp: $\mathbf{A} \rightarrow \exp(\mathbf{A})$, $\mathbf{A} \in$ sl(2, ℝ), and $\exp(\mathbf{A}) \in$ SL(2, ℝ) (Figure 4). For $\mathbf{A} \in$ sl(2, ℝ), Tr($\mathbf{A}$) = 0, and from det(exp($\mathbf{A}$)) = exp(Tr($\mathbf{A}$)) follows det(exp($\mathbf{A}$)) = 1. That is, exp($\mathbf{A}$) $\in$ SL(2, ℝ) because SL(2, ℝ) consists of matrices with determinant =1. Due to det(exp($\mathbf{A}$)) $\neq$ 0, each matrix of the group has its inverse. Exponents of the basis elements $\mathbf{S}_i$ are $\exp(t\mathbf{S}_1) = \begin{pmatrix} \exp(t) & \\ & \exp(-t) \end{pmatrix}$, $\exp(t\mathbf{S}_0) = \begin{pmatrix} \cos t & \sin t \\ -\sin t & \cos t \end{pmatrix}$, $\exp(t\mathbf{S}_2) = \begin{pmatrix} \cosh t & \sinh t \\ \sinh t & \cosh t \end{pmatrix}$ [14–17].

SL(2, ℝ) is a noncommutative multiplicative group. Noncommutativity of the group elements compositions may reflect nonreversibility of metabolic pathways, biochemical reactions, and so forth. The commutator [, ] gives an effective approximation of the noncommutative operations of the group. If the group elements are curves on $\mathbf{G} \in$ SL(2, ℝ), composition of two elements of $\mathbf{G}$, $\mathbf{gf}$ (left translation by $\mathbf{g}$ or multiplication from the left) will transform the curve $\mathbf{f}$ to the curve $\mathbf{gf}$. Right translation by $\mathbf{g}^{-1}$, $\mathbf{gfg}^{-1}$ will not move $\mathbf{gf}$ right backwards to the place coincident to the initial trajectory $\mathbf{f}$. $\mathbf{f} \neq \mathbf{gfg}^{-1}$. ($\mathbf{fg} \neq \mathbf{gf}$). Lie bracket substitutes the gaps, when results of the group operations are interpreted through the elements of the algebra.

For example, composition of two one-parameter subgroups of diffeomorphisms $\mathbf{g}^t\mathbf{f}^t$ (action of $\mathbf{g}^t$ on $\mathbf{f}^t$ from the left) displaces the initial integral curve $\mathbf{f}^t$ related to the tangent vector (let it be vector $\mathbf{A}$) at identity point. Because $\mathbf{g}^t$ and $\mathbf{f}^t$ do not commute, there is a difference in the results of the actions of $\mathbf{g}^t$ and $\mathbf{f}^t$, depending on the order of the composition: $\mathbf{g}^t\mathbf{f}^t - \mathbf{f}^t\mathbf{g}^t \neq 0$. The "gap" between two pathways $\mathbf{g}^t\mathbf{f}^t$ and $\mathbf{f}^t\mathbf{g}^t$ can be approximated by the Lie bracket of the derivations of $\mathbf{f}^t$ and $\mathbf{g}^t$ corresponding to the tangent vectors $\mathbf{A}$ and $\mathbf{B}$, respectively:

$$[\mathbf{A}, \mathbf{B}] = \mathbf{AB} - \mathbf{BA} = \lim t \longrightarrow 0 \left(\frac{t_2}{2}\right)\left(\mathbf{f}^t\mathbf{g}^t - \mathbf{g}^t\mathbf{f}^t\right). \quad (3)$$

In the small neighborhood exponent mapping transforms the structure of the Lie algebra into the structure of the group, or, more precisely, transforms an additive function and bracket operation of the algebra into multiplicative operation of the group

$$\begin{aligned}\mathbf{g}^t\mathbf{f}^t &= \exp(t\mathbf{A})\exp(t\mathbf{B}) \\ &= \exp\left[t(\mathbf{A} + \mathbf{B}) + \frac{t_2}{2}[\mathbf{A}, \mathbf{B}] + O(t3)\right].\end{aligned} \quad (4)$$

Basis elements $\mathbf{S}_0$, $\mathbf{S}_1$, and $\mathbf{S}_2$ have special for biologic applications properties related to their matrix structure. In the standard basis $\mathbf{e}_1 = \begin{bmatrix} 1 \\ 0 \end{bmatrix}$, $\mathbf{e}_2 = \begin{bmatrix} 0 \\ 1 \end{bmatrix}$ only operator $\mathbf{S}_1$ has diagonal matrix form $\mathbf{S}_1 = \begin{pmatrix} 1 & \\ & -1 \end{pmatrix}$. It has two one-dimensional *invariant* subspaces $L_1$ and $L_2$ direct sum of which $L_1 \oplus L_2 = L$ is the entire two-dimensional space $L$. $L_1$ is invariant under $\mathbf{S}_1$ means $\forall \mathbf{x} \in L_1 \rightarrow \mathbf{S}_1\mathbf{x} \in L_1$, ($\mathbf{x} \neq 0$). Vectors from $L_1$ are transformed by $\mathbf{S}_1$ into the vectors of the same subspace $L_1$. So operator action on the elements of invariant subspace leaves elements in the same subspace. This is also true for $L_2$, because $L_2$ is also an invariant subspace.

Vectors of the invariant subspace are *eigenvectors*, and each value of the matrix diagonal element is an *eigenvalue* of the eigenvectors of an eigenspace (invariant subspace). Real eigenvalue $+n$ gives $n$-times elongation of the applied vectors. If sign of eigenvalue is negative, it changes vectors directions to the opposite. Basis vectors $\mathbf{e}_1, \mathbf{e}_2$ are eigenvectors of $L_1$ and $L_2$, respectively, with eigenvalues $+1$ and $-1$. $\mathbf{S}_1$ will leave vectors of $L_1$ unchanged, because of the $+1$ eigenvalue and transform vectors of $L_2$ to the opposite (reflection) due to negative eigenvalue sign. Applied to the basis vectors (eigenvectors), $\mathbf{S}_1$ will transform $\mathbf{e}_1$ to the same vector and change direction of $\mathbf{e}_2$ to the opposite. Because $\mathbf{S}_1$ divides $L$ on two invariant subspaces, $\mathbf{S}_1$ can be considered as a direct sum of two operators $\mathbf{S}_1 = \mathbf{S}_1' \oplus \mathbf{S}_1''$ acting separately on the corresponding subspaces $L_1$ and $L_2$, so that, if $\mathbf{e} = \mathbf{e}_1 + \mathbf{e}_2$, $\mathbf{S}_1\mathbf{e} = \mathbf{S}_1'\mathbf{e}_1 + \mathbf{S}_1''\mathbf{e}_2$.

Biological meaning of invariant spaces suggests that some parts of the system have autonomous regulations. The term subsystem implies separable structure, which found its functional correlates in invariant subspaces. The existence of regulatory mechanisms (operators) acting separately on subsystems makes the subsystems anatomically distinguishable and functionally closed. Eigenvectors of invariant spaces can be interpreted as some conditions of the system, and eigenvalues—as some commands how to change the current condition's magnitude or direction, not their qualitative characteristics. Positive eigenvalues increase existing functional activity. Negative eigenvalues invert the process to opposite.

In the basis $\{\mathbf{e}_1, \mathbf{e}_2\}$ the operator $\mathbf{S}_2$, responsible for positive feedback, has matrix form $\mathbf{S}_2 = \begin{pmatrix} & 1 \\ 1 & \end{pmatrix}$. Its eigenvectors $\mathbf{f}_1 = \begin{bmatrix} 1 \\ 1 \end{bmatrix}$, $\mathbf{f}_2 = \begin{bmatrix} 1 \\ -1 \end{bmatrix}$ have $\pm 1$ eigenvalues (obtained from characteristic polynomial equation), so that $S_2$ can be transformed to the diagonal form by *similarity transformations* using orthogonal matrix $U$, $S_2' = U^{-1}S_2U$. $S_2'$ belongs to the same class of equivalence as the matrix $S_2$. This transformation changes the initial coordinate system $e = \{e_1, e_2\}$ to $f = \{f_1, f_2\}$ which corresponds to the eigenspaces $L_1'$, $L_2'$. The new coordinates could be obtained either by the rotation of coordinate axes clockwise on $45°$, or just by the rotations of the plane without coordinates (Figures 4 and 5). Invariant subspaces $L_1'$ and $L_2'$ are two lines inclined by $45°$ to the main coordinate axes (Figure 4).

Together operators $S_1$ and $S_2$ divide $L$ on four one-dimensional subspaces: $L_1 = \{ae_1 \mid a \in \mathbb{R}\}$, $L_2 = \{be_2 \mid b \in \mathbb{R}\}$, $L_1' = \{cf_1 \mid c \in \mathbb{R}\}$, and $L_2' = \{df_2 \mid d \in \mathbb{R}\}$. $S_1$ and $S_2$ transform points of $L$ living vectors from $L_1$, $L_2$ and $L_1'$, $L_2'$, in the same subspaces $S_1 a e_1 = a e_1$, $S_1 b e_2 = -b e_2$, $S_2 c f_1 = c f_1$, and $S_2 d f_2 = d f_2$. Although $S_2 = \begin{pmatrix} & 1 \\ 1 & \end{pmatrix}$ is transformed to the same diagonal form as $S_1 = \begin{pmatrix} 1 & \\ & -1 \end{pmatrix}$, $S_1$ and $S_2'$ are different objects, because similarity transformations change just the matrix forms of the operator $S_2$, which is different from the operator $S_1$. Unlike $S_1$ and $S_2$, $S_0 = \begin{pmatrix} & 1 \\ -1 & \end{pmatrix}$ is irreducible to one-dimensional invariant subspaces, because it has two complex eigenvalues $\pm i$. $S_0$ endows $L$ with the complex structure, which can be represented on the plane $\mathbb{R}^2$ of real variables. Thus, the only invariant subspace for $\mathbb{C}$ is $L$ itself, which is a two-dimensional space over $\mathbb{R}$. The operator $S_0$ just rotates vectors from $L$ leaving the vectors lengths unchanged.

If a smooth curve $g = g(t)$ on $L$ is given as a prototype of some biochemical, metabolic, or physiologic transformations, the tangent vectors to the curve at each point will give linear approximations of the real processes. The value of the tangent vector to some point on the curve is the velocity at which the curve is coming through this point. This is how matrices of $sl(2, \mathbb{R})$, used in ODE, create vector fields or fields of forces on $L = \mathbb{R}^2$, the spaces of biologic variables. The smooth curves on $\mathbb{R}^2$ (or $SL(2, \mathbb{R})$) are obtained as solutions of ODE. In other words, each (operator) matrix $(A, B, C)$ in ODE $(dx/dt = Ax, dx/dt = Bx, dx/dt = Cx)$ creates its own vector field by assigning to each point of the space of variables $V = (x_i, y_i)$ a vector (derivative), which is a velocity of the curve $g(t)$ coming through this point. A Lie algebra is a derivation of a Lie group, thus at each point $p$ of the group there is a tangent space $T_p$ of the elements of a Lie algebra. Each element of tangent space $T_p$ is a matrix corresponding to the vector field. Hence, every element of the algebra has a corresponding vector field. The value of this field, related, for example, to the matrix $A$ ($A : \mathbb{R}^2 \to \mathbb{R}^2$, $A \in sl(2, \mathbb{R})$) at arbitrary point $x \in \mathbb{R}^2$ is defined as $T_A(x) = Ax$. Matrix $A$ or related to $A$ vector field $T_A$ determines evolution of $x$ along *integral curves* $x(t) = \exp(tA)x_0$, which are solutions of the differential equation $dx/dt = Ax$.

Obtained integral curves lie on the surface $SL(2, \mathbb{R})$.

Correspondence between matrices of the algebra and vector fields tells us that matrix structure determines specific for each matrix physical (physiological) action.

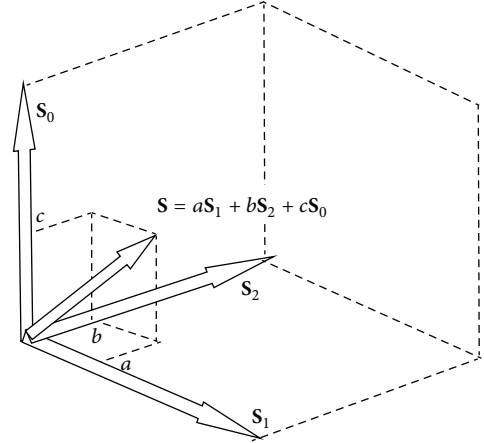

FIGURE 7: The basis matrices $S_0$, $S_1$, and $S_2$ of $sl(2, \mathbb{R})$ represent corresponding regulatory elements of a biologic system. Integrated functional unit $S$ is a linear combination of the basis.

In the standard basis $e_1 = \begin{bmatrix} 1 \\ 0 \end{bmatrix}$, $e_2 = \begin{bmatrix} 0 \\ 1 \end{bmatrix}$, the values of the vector fields, related to $S_0$, $S_1$, and $S_2$ at some point $(u, v) \in L$ are

$$T_{S_0} = \begin{pmatrix} & 1 \\ -1 & \end{pmatrix} \begin{bmatrix} u \\ v \end{bmatrix} = \begin{bmatrix} v \\ -u \end{bmatrix},$$

$$T_{S_1} = \begin{pmatrix} 1 & \\ & -1 \end{pmatrix} \begin{bmatrix} u \\ v \end{bmatrix} = \begin{bmatrix} u \\ -v \end{bmatrix}, \quad (5)$$

$$T_{S_2} = \begin{pmatrix} & 1 \\ 1 & \end{pmatrix} \begin{bmatrix} u \\ v \end{bmatrix} = \begin{bmatrix} v \\ u \end{bmatrix}.$$

Written in the differential form $T_{S_0} = (v\partial u - u\partial v)$, $T_{S_1} = (u\partial u - v\partial v)$, and $T_{S_2} = (v\partial u + u\partial v)$ these vector fields will serve as *infinitesimal generators* of the group action on the space of biologic variables $L$. It can be shown that $[T_{S_i}, T_{S_j}] = T_{[S_i, S_j]}$; that is, vector fields with the basis elements $\{T_{S_i}\}$ form a Lie algebra $\mathfrak{a}$ isomorphic to the algebra $\mathfrak{g} \in sl(2, \mathbb{R})$.

Because $\mathfrak{g} = sl(2, \mathbb{R})$ is a vector space over $\mathbb{R}$, linear combinations of $S_i$ with real coefficients $\{a, b, c\}$ will give matrices $S$ from $\mathfrak{g}$; $S = aS_0 + bS_1 + cS_2$. They will correspond to the integrated vector field generated by $S$ on $L$ (Figure 7).

Being a linear combination of $S_i$, $S$ assumes a simultaneous action of three basis elements, so that each condition of two-element system is determined by the states of the three basis regulatory subsystems, functioning together. These subsystems are positive feedback and negative feedback and reciprocal links. They link morphologic elements in a whole functional structure. The three basis elements of the Lie algebra $\mathfrak{g} = sl(2, \mathbb{R})$ are "infinitesimal" analogs of positive feedback and negative feedback and reciprocal links.

Points $p(x, y)$ of the space of two variables $x \in X$, $y \in Y$ determine all possible conditions $(x, y) \in X \times Y = L$ of the system. Variables change their values in time, and these changes for $x$ and $y$ are points on the curves $\gamma : t \to X$, $\varphi : t \to Y$. The characteristics of the curves are determined by the matrix $S$, which also determines functional relations between $x$ and $y$. Due to nonsingularity of $S$, these relations can be expressed by a function $y = F(x)$.

$F$ determines a family of trajectories on $X \times Y$. Any of them refers the process to some initial condition $(x_0, y_0)$. The phase trajectories lie on the 2-dimensional surface of two variables. They are projections of the integral curves $x(t) = \exp(t\lambda_1)x$, $y(t) = \exp(t\lambda_2)y$, which are solutions of the system of the two ordinary differential equations $dx/dt = \Phi_1(x, y)$, $dy/dt = \Phi_2(x, y)$, where $\Phi_1(x, y) = ax + by$, $\Phi_2(x, y) = cx + dy$, $\mathbf{S} = \begin{pmatrix} a & b \\ c & d \end{pmatrix}$, $\mathbf{S} \in \mathrm{sl}(2, \mathbb{R})$. Transforming $\mathbf{S}$ to one of the Jordan forms we obtain eigenvalues $\lambda_1$ and $\lambda_2$ and eigenvectors as new variables $x'$, $y'$. Integral curves $x'(t)$ and $y'(t)$ are one-parameter group of diffeomorphisms $\alpha^t(x', y') : \mathbb{R} \to (\exp(t\lambda_1)x', \exp(t\lambda_2)y')$. Each function $\exp(t\lambda_i)$ transforms corresponding variable from $X \times Y$, satisfying group properties, hence determining two-directional process. A curve $\alpha^t(x, y)$ is a direct product of two one-parameter groups of diffeomorphisms $\exp(t\lambda_1)$, $\exp(t\lambda_2)$ on $X$ and $Y$.

The Lie algebra $\mathfrak{g}$ can act not only in the space of biologic variables $L$, but also on its own elements and elements of the Lie group $\mathbf{G}$. Thus we can apply some regulatory structures to the primary regulatory elements themselves, considering it as the next hierarchical level of regulations.

Using matrices $\mathrm{sl}(2, \mathbb{R})$ one can create vector fields on elements of $\mathrm{SL}(2, \mathbb{R})$. For the space of matrix elements $\mathbf{G} = \mathrm{SL}(2, \mathbb{R})$ and fixed $2 \times 2$ real matrix $\mathbf{A} \in \mathfrak{g}$, $\mathfrak{g} = \mathrm{sl}(2, \mathbb{R})$, let $\mathbf{L_A}$ be a vector field on $\mathbf{G}$ generated by $\mathbf{A}$ whose values on each point (matrix) $\mathbf{g} \in \mathbf{G}$ are

$$\mathbf{L_A}(\mathbf{g}) = \mathbf{g}\mathbf{A}. \tag{6}$$

We relate the defined vector field to the ODE $d\mathbf{g}/dt = \mathbf{g}\mathbf{A}$, solution of which is integral curves $\mathbf{g}(t)$ on $G$:

$$\mathbf{g}(t) = \mathbf{g}_0 \exp(t\mathbf{A}). \tag{7}$$

The curve $\mathbf{g}(t)$ is a result of an action of the one-parameter subgroup of diffeomorphisms $(\exp(t\mathbf{A}))$ generated by $\mathbf{A}$ on element $g$ of the group $G$. The vector $\mathbf{L_A}(\mathbf{g}) = \mathbf{g}\mathbf{A}$ is a tangent vector to the surface $\mathbf{G}$ at the point $\mathbf{g} \in \mathbf{G}$, and it is a velocity at which $\mathbf{g}\exp(t\mathbf{A})$ is coming through that point $g$. The points of the curve created by the vector field $\mathbf{L_A}(\mathbf{g})$ are obtained by multiplying the matrix exponent $\exp(t\mathbf{A})$ on $\mathbf{g}$.

By applying matrix $\mathbf{f} \in \mathbf{G}$ to the vector field from the left an integral curve is transformed to the new point $\mathbf{fg}$:

$$\mathbf{f}\mathbf{L_A}(\mathbf{g}) = \mathbf{L_A}(\mathbf{fg}). \tag{8}$$

This is called *left-invariance* of the field. Because $\mathbf{A}$ is an element of the Lie algebra $\mathrm{sl}(2, \mathbb{R})$, $\mathbf{A}$ lies on the tangent surface to the group $\mathrm{SL}(2, \mathbb{R})$ at the identity point, and due to the left invariance, $\mathbf{L_A}$ creates vector fields related to $\mathbf{A}$ on the entire group $\mathrm{SL}(2, \mathbb{R})$. There are also commutation rules for vector fields created from the elements of the algebra

$$[\mathbf{L_A}, \mathbf{L_B}] = \mathbf{L_{[A,B]}}, \tag{9}$$

so that vector fields on $G$ have the structure of a Lie algebra $\mathrm{sl}(2, \mathbb{R})$.

If $\mathbf{A}$, $\mathbf{B}$, and $\mathbf{C}$ are the basis vectors of $\mathfrak{g}$, related vector fields $\mathbf{L_A}$, $\mathbf{L_B}$, and $\mathbf{L_C}$ are linearly independent, and at each point $\mathbf{g} \in \mathbf{G}$ they present the three basis elements $\mathbf{L_A}(\mathbf{g})$, $\mathbf{L_B}(\mathbf{g})$, and $\mathbf{L_C}(\mathbf{g})$ of three-dimensional vector space $T_g$ tangent to $\mathbf{G}$ at $\mathbf{g} \in \mathbf{G}$ [18, 19].

## 5. Actions of the $\mathrm{sl}(2, \mathbb{R})$ Algebra on the Space of Biologic Variables and on Its Own Elements

It is proposed that the operators (matrices) $\mathbf{S}_0$, $\mathbf{S}_1$, and $\mathbf{S}_2$ represent regulatory basis of the two elements systems. The $2 \times 2$ matrices over $\mathbb{R}$ transform points of a two-dimensional *carrier space* of biologic variables. For the genetically fixed regulatory mechanisms, which were assumed to be positive and negative feedback and reciprocal links, the corresponding morphological structures should also be fixed, such as, for example, central and peripheral endocrine glands, realizing negative feedback regulatory "algorithm," and antipodal structural elements, which are ready to release counter acting substances—insulin versus glucagon, vWF (von Willebrand Factor) versus anticoagulant proteins C and so on. Although each of the three basis regulatory elements is acting on two element structures and links only two variables, it is not evident that substantially different mechanisms simultaneously or intermittently can be applied to the same two variables. It seems unconvincing that, for example, central and peripheral endocrine glands, normally operating through the feedback loops, are also under the direct control of reciprocal links and positive feedback mechanisms. Negative feedback implies hierarchy; that is, corresponding variables should always be on different functional levels. On the contrary, the reciprocity links variables, which are on the same line of functional capabilities, so they are on the same hierarchical level. It seems to be also true for positive feedback in the sense that each of the basis regulatory elements fixes some pairs of variables becoming a basis for corresponding two-dimensional subsystems. It is assumed, that each of the operators $\mathbf{S}_0$, $\mathbf{S}_1$, and $\mathbf{S}_2$ has corresponding two-dimensional space $L_{S_i}$, and the sum of these spaces $L_S = L_{S_0} + L_{S_1} + L_{S_2}$ is a six-dimensional space of variables $L_S = \mathbb{R}^6$. On the other hand, $\mathbf{S}_0 \cap \mathbf{S}_2 \ne 0$, and, due to correspondence between $L_{S_i}$ and $\mathbf{S}_i$, it follows that $L_{S_0} \cap L_{S_2} \ne 0$, and the dimension of $L_S$ is less than six. Nevertheless, we will consider $L_S$ as a direct sum of $L_{S_i}$, $L_S = L_{S_0} \oplus L_{S_1} \oplus L_{S_2}$, which is a six-dimensional space, because $\mathbf{S}_i$ as the basis elements of a Lie algebra $\mathfrak{g}$ are linearly independent. Additional, physiological reason for $L_S = \mathbb{R}^6$ is that each $\mathbf{S}_i$ is an "irreducible" unit that operates on "its own" two-dimensional subspace $L_{S_i} = \mathbb{R}^2$.

From this it follows that the "integrated" operator $\mathcal{S}$ on $L_S$ is a direct sum of $\mathbf{S}_0$, $\mathbf{S}_1$, $\mathbf{S}_2$, $\mathcal{S} = \mathbf{S}_0 \oplus \mathbf{S}_1 \oplus \mathbf{S}_2$. In the standard basis $\{\mathbf{e}_1, \mathbf{e}_2, \mathbf{e}_3, \mathbf{e}_4, \mathbf{e}_5, \mathbf{e}_6\}$ of $L_S = \mathbb{R}^6$ the matrix of $\mathcal{S}$ has block-diagonal form and the blocks are matrices $\mathbf{S}_0$, $\mathbf{S}_1$, and $\mathbf{S}_2$ (Figure 8).

If pairs of variables are fixed for each basis regulatory mechanism ($\mathbf{S}_i$ matrix), then the six-space of biologic variables could be stemmed by three pairs of antipodal, reciprocally related, variables linked by negative and positive feedback loops. This hypothetical network is demonstrated in Figure 9. There are examples of the mechanisms maintaining rheology of the blood, which support this functional structure. According to the current understanding of these mechanisms fibrinolysis and coagulation are regulated by positive and negative feedback loops acting simultaneously on each of these components [20]. In order to comprise closed regulatory structure, elements of fibrinolytic and coagulation cascades are supposed to be connected by reciprocal links. That is, the set of either $X$, $X'$, $X''$, or $Y$, $Y'$, $Y''$ variables shown in Figure 9 can be considered as a clot formation or clot degradation subsystems linked by $S_1$ (reciprocal) types of relations. Physiologic meaning of $S_1$ (= reciprocal links) is that, if, in the normal conditions, coagulation begins to be a prevailing component, it will stimulate a fibrinolytic activity and simultaneously slow down the clot formation processes. Another example, demonstrating possibility for each subsystem to have different types of simultaneously acting regulatory loops with other subsystems is a hypothalamic-hypophyseal-thyroid axis. A thyroid gland linked by negative feedback loops with hypothalamus, at the same time, is connected by reciprocal relations with parathyroids in the regulation of calcium metabolism. Production of calcitonin by C-cells of thyroid glands is also regulated by negative feedback through the concentration of calcium in the blood. On hypophyseal level ultrashort feedback loops are found as additional to the negative feedback with thyroids, means of regulation of thyroid stimulating hormone (TSH) secretion [21, 22]. These data support the existence of different regulatory mechanisms applied to the same anatomical structure (or subsystem).

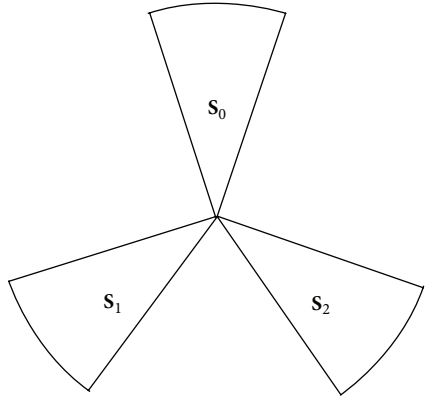

Figure 8: Direct sum of $S_i$ corresponds to $\mathbb{R}^6$ space of biologic variables.

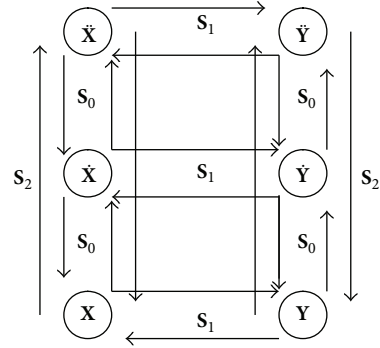

Figure 9: Schematic representation of the functional regulatory structure of a biologic system.

The graph shown in Figure 9 is a single regulatory unit. It is closed and contains minimal number of links corresponding to the three basis operators. If only one variable is considered in a stem, then functional unit including all three regulatory components will correspond to four-dimensional carrier space. It is clearly seen for the pairs of variables and links that originated from only one ($x$ or $y$) variable (Figure 9).

If $S_i$ are given as the basis for $sl(2, \mathbb{R})$, the linear span $\langle S_0, S_1, S_2 \rangle$ is a three-dimensional linear space of elements $S = aS_0 + bS_1 + cS_2$ over $\mathbb{R}$, $S \in sl(2, \mathbb{R})$. In this case the carrier space remains two-dimensional, because $S$ is $2 \times 2$ (traceless) matrix. $S$ represents the action on two-dimensional carrier space embedded in the six-dimensional space.

The algebra $\mathfrak{g}$ itself, the $\mathbb{R}^3$ space, can be considered as a carrier space for some transformations $F : \mathfrak{g} \to \mathfrak{g}$. In fact, *adjoint* transformations of $sl(2, \mathbb{R})$ elements (discussed further in the text) have the same algebra structure (it is $sl(2, \mathbb{R})$) as transformations of biologic variables.

As soon as $\{S_i\}$ fixes the basis elements of the space of biologic variables $L$, the $\mathbb{R}^6$ spaces of biologic variables and the $\mathbb{R}^3$ space of transformations, which are the space of the algebra elements spanned by $S_0$, $S_1$, and $S_2$, are related.

There is still an ambiguity in determining the dimension of a carrier space of biologic variables. Since $S$ is a linear combination of basis transformations $S_0$, $S_1$, and $S_2$, composition of these elements gives integrated matrix structure that predisposes simultaneous actions of $S_i$. Therefore, the all three basis elements $S_0$, $S_1$, and $S_2$ as an integrated unit, should be linked through the same variable or relate the same two variables. On the other hand, each $S_i$, as emphasized before, acts on its own, corresponding to $S_i$, space of variables $L_{S_i}$, thus making the whole space not two, or even four-dimensional, but, because of symmetry of reciprocal links, a six-dimensional (Figure 9).

To clarify this uncertainty two separate time intervals in the course of phylogenetic development should be considered, early stage, consisting of rudimental functions, and advanced stage, when mature structures have already been formed. In the first (early) stage of functional isotropy conditions of the system were, for instance, points of $n$-dimensional space $V$. In this stage there was no means to regulate each of $n$ parameters separately, because $V$ is the only invariant space, and it was irreducible to the invariant subspaces of lesser dimensions. For simplicity, $n = 2$, so we can describe internal regulatory structure through the phase trajectories showing changes in the system's conditions. In chosen for $V$ basis $\{l, m\}$ the matrices of transformations of $V$ will have common form $\mathbf{M} = \begin{pmatrix} a & b \\ c & d \end{pmatrix}$, where $\mathbf{M}$ is a SL(2, $\mathbb{R}$) group element. Because $\mathbf{M}$ reflects regulatory structure of mature (not growing) biologic objects, $\det(\mathbf{M}) = 1$, meaning $\mathbf{M}$ is volume preserving transformations, preventing systems from being expanded or collapsed. Unit determinant $\det(\mathbf{M}) = ad - bc = 1$ restricts 4-space to the 3-space of transformations.

Depending on the values of the matrix entries, $\mathbf{M}$ can be transformed to the simpler forms. There are three possibilities determined by the roots $\lambda_1, \lambda_2$ of a characteristic polynomial equation:

$$\lambda^2 - \text{Tr}(\mathbf{M})\lambda + \det(\mathbf{M}) = 0,$$
$$\lambda_{1,2} = \frac{1}{2}\text{Tr}(\mathbf{M}) \pm \frac{1}{2}\sqrt{\text{Tr}^2(\mathbf{M}) - 4\det(\mathbf{M})}, \quad (10)$$

$\text{Tr}^2(\mathbf{M}) = (a+d)^2 < 4$. $\lambda_{1,2}$ are complex numbers, $m \pm m$, $\mathbf{M}' = \begin{pmatrix} m & n \\ -n & m \end{pmatrix}$; then $\mathbf{M}$ is an *elliptic* element.

Consider that $\text{Tr}^2(\mathbf{M}) = (a+d)^2 = 4$. $\lambda_{1,2}$ are +1 or −1, $\mathbf{M}'' = \begin{pmatrix} 1 & 1 \\ & 1 \end{pmatrix}$, or $\mathbf{M}'' = \begin{pmatrix} -1 & 1 \\ & -1 \end{pmatrix}$, where $\mathbf{M}$ is a *parabolic* element.

Consider that $\text{Tr}^2(\mathbf{M}) = (a+d)^2 > 4$. $\lambda_{1,2}$ are different real numbers, $\mathbf{M}''' = \begin{pmatrix} p & \\ & q \end{pmatrix}$, where $\mathbf{M}$ is a *hyperbolic* element.

Similarity transformations resulted in any of $\mathbf{M}', \mathbf{M}'', \mathbf{M}'''$ change the initial basis in which $\mathbf{M}$ was expressed. And, in these new bases and related matrix forms $\mathbf{M}', \mathbf{M}'', \mathbf{M}'''$, $V$ may have one-dimensional ($\mathbf{M}'', \mathbf{M}'''$) or two-dimensional ($\mathbf{M}'$) invariant subspaces. Importance of this for morphological and functional development is that the initial space becomes reducible to the subspaces of lesser dimensions. These subspaces can be maintained and developed further as autonomous systems. Thus functional properties of operators of $\text{SL}(2,\mathbb{R})$ implies the existence of three types of regulatory structures. In biological terms, regulatory structure of the system, imitating the group structure of $\text{SL}(2,\mathbb{R})$ gives three autonomous branches of transformations from the primary stem of not yet separable functions.

During phylogenetic development the changes in the bases of the carrier spaces eventually are becoming fixed genetically in new anatomical structures having autonomous features corresponding to invariant subspaces. Primary morphological and functional stem has been split on two distinguishable branches with respect to the invariant subspaces. Thus new anatomical structures have become supported by steady regulatory mechanisms. They were assumed to be negative feedback, positive feedback, and reciprocal links.

The three basis operators comprising elements of the Lie algebra $\mathfrak{g} = \text{sl}(2,\mathbb{R})$ are linear analogues of these phylogenetic mechanisms.

Matrix $\mathbf{S} = \begin{pmatrix} a & b+c \\ b-c & -a \end{pmatrix}$, $a, b, c \in \mathbb{R}$, of transformations on $L = \mathbb{R}^2$ is a linear combination of the elements of $\{\mathbf{S}_0, \mathbf{S}_1, \mathbf{S}_2\}$ basis. The "integrated" matrix $\mathbf{S}$ can be transformed to the diagonal (Jordan) form and be expressed through the basis of eigenvectors. To obtain this matrix, we have to find the roots of the characteristic equation $\det(\mathbf{S} - \lambda \mathbf{E}) = 0$. The roots $\lambda_1, \lambda_2$ of the characteristic polynomial equation are eigenvalues of the vectors of invariant subspaces: $\det(\mathbf{S}-\lambda \mathbf{E}) = -(a-\lambda)(a+\lambda) - (b^2-c^2) = 0$. Consider $\lambda_{1,2} = \pm\sqrt{(a^2+b^2-c^2)}$. Depending on the values of $a, b, c$, there are three possibilities: $\lambda_{1,2}$ can be real, zero, or complex numbers. For a more demonstrative picture, $\mathbf{S}$ can be viewed as an operator of $d\mathbf{x}/dt = \mathbf{Sx}$.

*Scenario 1.* $\lambda_{1,2}$ are real, if $(a^2 + b^2 - c^2) > 0$; $\lambda_{1,2} = \pm p$; $\mathbf{S}_1' = \begin{pmatrix} p & \\ & -p \end{pmatrix}$. Therefore, operator $\mathbf{S}$ may have two invariant one-dimensional subspaces containing eigenvectors with real eigenvalues. Formally, that means that $\mathbf{S}$ can be transformed to the diagonal form by *similarity transformations* $\mathbf{S}_1' = \mathbf{T}^{-1}\mathbf{ST}$, where $\mathbf{T}$ for this case is an idempotent orthogonal matrix $\mathbf{T} = \begin{pmatrix} 1 & 1 \\ & -1 \end{pmatrix}$. Eigenvectors for the operator $\mathbf{S}$ are $\varphi_1 = \begin{pmatrix} 1 \\ 0 \end{pmatrix}$, $\varphi_2 = \begin{pmatrix} 1 \\ -1 \end{pmatrix}$. In new coordinates $\psi_1 = \varphi_1$, $\psi_2 = \varphi_1 - \varphi_2$, and eigenvectors are orthogonal and can be taken as coordinate axes for the whole space $L$ (Figures 4 and 5).

*Scenario 2.* $\lambda_{1,2}$ are complex numbers, $(a^2 + b^2 - c^2) < 0$; $\mathbf{S}_0' = \begin{pmatrix} & m \\ -m & \end{pmatrix}$. The entire two-dimensional space $L$ is an invariant space (Figure 3).

*Scenario 3.* $\lambda_{1,2} = 0$, $(a^2 + b^2 - c^2) = 0$, $\det \mathbf{S} = 0$, decays on two cases with the singular matrices $\mathbf{S}' = 0$, or $\mathbf{S}'_{\text{nil}} = \begin{pmatrix} 0 & 1 \\ 0 & 0 \end{pmatrix}$. In the first case all the points of the space are stationary points. The second one, related to the nilpotent operator, is a marginal situation for the improper node when two equal positive roots aimed to zero. Its eigenspaces are stationary points of horizontal coordinate axis. Nilpotent operator $\mathbf{S}'_{\text{nil}}$ has only one invariant one-dimensional subspace. All these scenarios are examples of similarity transformations $\mathbf{S}'_i = \mathbf{M}^{-1}\mathbf{SM}$, $\mathbf{S} \approx \mathbf{S}_i$.

Each value of a physiologic variable reflects some condition of a system, which is the result of interactions of comprising the system subsystems. Superposition of, for example, $n$ subsystems organized in 2-element structures is equivalent to the interactions of $2n$ or $n+1$ variables, depending on whether subsystems have common elements. It is postulated that for two-element subsystems only three types of regulations, negative feedback ($\mathbf{S}_0$), positive feedback ($\mathbf{S}_2$), and reciprocal links ($\mathbf{S}_1$), are genetically and anatomically fixed. Other forms are "linear combinations" of these three.

A system described by $n$ variables can be reduced to the group of simpler two-component structures; it is easier to trace a behavior of two-element system as well. Thus, we consider two-dimensional space $L$ as a space of pairs $(x, y) \in X \times Y$ of biologic variables $X, Y$ of a two-element system. For example, these components are $\{X = $ clot formation system, $Y = $ clot degradation system$\}$, $\{X = $ insulin-dependent system lowering the glucose of the blood, $Y = $ glucagon-dependent system facilitating glucose production and increase of its concentration in the blood$\}$, $\{X = $ system, eliminating Na$^+$ from the blood (decreasing its concentration), $Y = $ system retaining Na$^+$ in the blood$\}$, and so forth.

Regulatory process keeping pairs of variables within optimal levels is determined by three operators $\mathbf{S}_0, \mathbf{S}_1, \mathbf{S}_2$, which are the basic functional components of a SYSTEM as a regulatory machine.

The following summarizes the main features, the "logic," of three-component regulatory system. Consider $L$ a linear two-dimensional space of biologic variables $(x)$ and $(y)$. Regulatory actions $\mathbf{g} \in \text{SL}(2,\mathbb{R})$ change conditions of the system according to some phase trajectories $F(x, y)$, which are projections of $\mathbf{g}(t)$ on $L$. Consider that $d\mathbf{g}/dt = \mathbf{S}$ at $t = 0$. $\mathbf{S}$ determines the vector field on $L$. $\mathbf{S} \in \text{sl}(2,\mathbb{R})$ is an arbitrary three-dimensional vector $\mathbf{S} = \mathbf{S}_0 + b\mathbf{S}_1 + c\mathbf{S}_2$ over $\mathbb{R}$, $\mathbf{S} \in \mathbb{R}^3$. In standard basis on $L\{\mathbf{e}_1, \mathbf{e}_2\}$ operator $\mathbf{S}$ has a form $\mathbf{S} = \begin{pmatrix} a & b+c \\ b-c & -a \end{pmatrix}$, $a, b, c \in \mathbb{R}$, which is a linear combination of the basis vectors $\mathbf{S}_0 = \begin{pmatrix} & 1 \\ -1 & \end{pmatrix}$, $\mathbf{S}_1 = \begin{pmatrix} 1 & \\ & -1 \end{pmatrix}$, $\mathbf{S}_2 = \begin{pmatrix} & 1 \\ 1 & \end{pmatrix}$ of the Lie algebra $\mathfrak{g}$, tangent to the space of transformations $SL(2, \mathbb{R})$. Basis vectors on $L\{e_1, e_2\}$ are considered as biologic variables $(x, y)$, so that the regulatory process $g \in SL(2, \mathbb{R})$ relating $x$ and $y$ is viewed according to the matrix $S$ tangent to $g$. $S$ (obtained, e.g., from experimental data and then linearized) is in general view, which does not show *regulatory structure* of $S$. There are three possible types of similar to $S$ simpler matrix forms. They are diagonal, symplectic, and nilpotent matrices. For example, $S$, because of the values of coefficients, is transformed to the diagonal form $S' = \begin{pmatrix} m \\ & -m \end{pmatrix} = m \begin{pmatrix} 1 \\ & -1 \end{pmatrix}$. According to that, initial basis $\{e_1, e_2\}$ changes, say, to $\{f_1, f_2\}$, where the basis elements $f_i$ are linear combinations of $e_i$. This means that in the new variables $x'$, $y'$ operator $S$ has the matrix form $S'$ so that the same regulatory process, but in terms of new variables, has become viewed as if consisting of two reciprocal (antipodal) subsystems. Relative to that, the new basis vectors will represent invariant subspaces; that is, only in a new variables $x'$, $y'$ it is possible to distinguish autonomous subsystems, regulated by $S$. If intrinsic characteristics of $S$ were related to $S'_0$ or $S'_{nil}$ structure, the new variables will be linked by negative feedback ($S'_0$), or some degenerative form ($S'_{nil}$), needed special consideration.

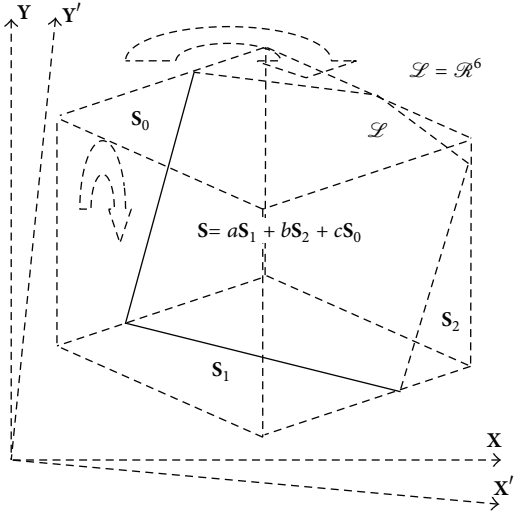

Figure 10: Similarity transformations $S_i = ASA^{-1}$ change initial basis of a carrier space $L$ as well as the basis of the space of transformations.

Similarity transformation $S \rightarrow S'$ transforms the basis $\{S_i\}$ to a new one $\{H_i\}$. Geometrically this transformation is related to the rotations of the three-dimensional surface $sl(2, \mathbb{R})$ to the point in which $S$, a two-dimensional surface of 4-space of linear transformations embedded in $\mathbb{R}^3$, coincides with or will be parallel to $S'_1$, $S'_0$, or $S'_{nil}$. The curve in $\mathbb{R}^3$ (or phase trajectory in $\mathbb{R}^2$) related to $S$ makes an additional movement in $\mathbb{R}^6$ in order to match with corresponding pair of variables (Figure 10).

It can be summarized by the following.

**Proposition 1.** *For each basis $\{e_1, e_2\}$ of a carrier linear space $L = \mathbb{R}^2$ there exist an orthonormal basis $\{H_0, H_1, H_2\}$ of transformation matrices $S \in \mathcal{L}$ on L, which can be transformed to the basis, having the view $S_0$, $S_1$, $S_2$, ($\{H_0, H_1, H_2\} \rightarrow \{S_0, S_1, S_2\}$). This transformation is coherent with the transformation of the basis elements $\{e_1, e_2\} \rightarrow \{f_1, f_2\}$ of the carrier space L.*

## 6. Automorphisms of Lie Algebra $sl(2, \mathbb{R})$ Imply the Existence of Uniformly Organized Regulatory Elements of Biologic System on Different Hierarchical Levels

Inner automorphisms of the group $SL(2, \mathbb{R})$, $I_a : G \rightarrow G$, $I_a = \mathbf{afa}^{-1}$, and automorphisms of the algebra $sl(2, \mathbb{R})$ $I_X : \mathfrak{g} \rightarrow \mathfrak{g}$, $I_X = \mathbf{YXY}^{-1}$ are not commutative operations, so $\mathbf{fa} \neq \mathbf{af}$ and for the algebra elements Lie bracket $[\mathbf{X, Y}] = \mathbf{XY} - \mathbf{YX}$ is a precise means to measure the "commutation gaps," which are elements of the algebra as well; $[\mathbf{X, Y}] \in \mathfrak{g}$.

The Lie bracket can be viewed in the other way, as linear transformations on elements of the algebra. By definition, *adjoint* action of $\mathbf{X}$ on algebra $\mathfrak{g}$, $\mathbf{ad_X}$, is a map, which with some fixed element of the algebra $\mathbf{X}$, sends elements $\mathbf{Y}$ to $[\mathbf{X, Y}]$. Thus $\mathbf{ad_X Y}[\mathbf{X, Y}] = \mathbf{XY} - \mathbf{YX}$. Each $\mathbf{X}$ has its adjoint representative in $\mathcal{F} : \mathfrak{g} \rightarrow \mathfrak{g}$; that is $\mathbf{X} \rightarrow \mathbf{ad_X} \in \mathcal{F}$. Consider that $[\mathbf{ad_X, ad_Y}] = \mathbf{ad_{[X,Y]}}$ follows from Jacobi identity, so $\mathbf{X} \rightarrow \mathbf{ad_X}$ is a *homomorphism* of $\mathfrak{g}$ and linear transformations $\mathcal{F}$ on $\mathfrak{g}$. The basis $\{S_0, S_1, S_2\}$ gives three $3 \times 3$ adjoint matrices of transformations on $\mathbb{R}^3 : \mathbf{ad_{S_0}} = \begin{pmatrix} & 2 \\ -2 & & 0 \end{pmatrix}$, $\mathbf{ad_{S_1}} = \begin{pmatrix} & & 2 \\ 0 & 2 & \end{pmatrix}$, $\mathbf{ad_{S_2}} = \begin{pmatrix} & & -2 \\ -2 & 0 & \end{pmatrix}$ (Empty spots of the 3x3 matrices $\mathbf{ad_S}$ are zeros), which form a basis of a Lie algebra of endomorphisms $\mathbf{End}(\mathfrak{g})$ on $\mathfrak{g}$.

Adjoint representation of the algebra are related to adjoint representation of the group through the exponent mapping $\exp: \mathbf{ad_X} \rightarrow \mathbf{Ad_g}$, and differentiating $\mathbf{Ad_g}$ returns it back to $\mathbf{ad_X}$, $d\mathbf{Ad_g}|_e = \mathbf{ad_X}$.

Unlike similarity transformations of elements of linear spaces, *inner automorphisms* of $SL(2, \mathbb{R})$, $I_a(\mathbf{f}) = \mathbf{afa}^{-1}$, are not precise: for example, $F_g(\mathbf{h}) = \mathbf{ghg}^{-1}$ transforms $\mathbf{h}$ to the different point $\mathbf{h}' \neq \mathbf{h}$. If elements of the algebra are to be transformed, for instance, $\mathbf{X} \in \mathfrak{g}$, the results of inner automorphisms $F_g(\mathbf{X}) = \mathbf{gXg}^{-1}$ will permute elements of the tangent space at the neutral element of the group, sending $\mathbf{X}$ to $\mathbf{Ad_g}(\mathbf{X})$, $\mathbf{X} \rightarrow \mathbf{Ad_g}(\mathbf{X}) \in \mathfrak{g}$. $\mathbf{Ad_g}$ is defined as a *derivation* of $F_g$ [18].

Restrictions of ($\mathbf{ad_{S_i}}$) operators on two-dimensional spaces give $2S_0$ and $\pm 2S_2$, which differ from the basis elements of $\mathfrak{g}$ $S_0$ and $S_2$ only by integer 2 and negative sign. Thus, the structure of the basis operators on $\mathbf{L}$ and $\mathbf{End}(\mathfrak{g})$, which is considered as the next, higher, and regulatory level, remains in general the same.

## 7. Metrics of the Space of Biologic Variables and the Space of Transformations

An interesting idea is to define quantitative characteristics of the functional elements of the system in the context of geometrical structure of the space of regulatory functions.

The *metric* gives such a measure of intrinsic properties of geometrical objects expressed through the lengths of vectors and norms of matrices. Another meaning of the metric is related to the characteristics of the spaces as geometrical surfaces in which matrices of transformations operate. The significance of metric for biologic objects is that it tells us what functional structure and regulatory characteristics a biologic system are preserved. The metric gives a natural way to find *invariants* of regulatory mechanisms.

In general, metric gives some quantitative measure to the geometrical objects. Formally, metric is a symmetrical bilinear form $\mathbf{Gr}: \mathbf{V} \times \mathbf{V} \to \mathbb{R}$. It relates pairs of vectors from $L$ to real numbers by the Gramm's matrices $\mathbf{Gr}$. In Euclidean space $\mathbb{R}^3$, a bilinear form is given by a scalar (inner) product of two vectors $\mathbf{x}$ and $\mathbf{y}$. For example, the square of a rectangle is a scalar product of its sides. In general, it can be written in the form $\mathbf{x g y} = \mathbf{g}(\mathbf{x}, \mathbf{y}) = g_{11} x^1 y^1 + g_{22} x^2 y^2 + g_{33} x^3 y^3$, where $\mathbf{g} \in \mathbf{Gr}$ is a metric matrix. If $\mathbf{g}$ is related to Euclidean space, $\mathbf{g} = \text{diag}(1, 1, 1)$, $g_{11} = g_{22} = g_{33} = 1$, then $\mathbf{g}(\mathbf{x}, \mathbf{y}) = x^1 y^1 + x^2 y^2 + x^3 y^3$. The length or the *norm* of the vector $\mathbf{x}$ is defined as a square root of the scalar product of the vector with itself $|\mathbf{x}| = \sqrt{\mathbf{g}(\mathbf{x}, \mathbf{x})}$. This gives the Pythagorean formula to calculate the lengths through the vector's coordinates $|\mathbf{x}| = \sqrt{(x^1)^2 + (x^2)^2 + (x^3)^2}$. This is an example of positive definite metric since all signs of the diagonal elements $\mathfrak{g}$ are positive numbers. The lengths of the vectors related to this metric are always $>0$.

If some of the signs of diagonal elements of a metric matrix are negative, metric is called *indefinite*, and the objects may have positive, negative, or zero lengths, depending on the values of components. For example, $\mathbf{g} = \text{diag}(-1, 1, 1)$ gives $|\mathbf{x}| = \sqrt{-(x^1)^2 + (x^2)^2 + (x^3)^2}$ and minus sign at the first component of the sum gives all three possibilities. The relativistic Minkowski space-time $\mathbb{R}_1^3$ has indefinite metric $\mathbf{g} = \text{diag}(-1, 1, 1, 1)$, which makes $\mathbb{R}_1^3$ a "curved" space.

By definition, $\mathbf{g} \in \mathbf{Gr}$ is a *metric tensor* of type $(0, 2)$, which determines a tensor field on a manifold $G = \text{SL}(2, \mathbb{R})$ by assigning to each point $\mathbf{p}$ of the manifold a scalar product $\mathbf{g}_p(\mathbf{X}, \mathbf{Y}), \mathbf{X}, \mathbf{Y} \in \mathfrak{g} = \text{sl}(2, \mathbb{R})$, on the tangent space $T_p$ to this point. If $\mathbf{S}_0$, $\mathbf{S}_1$, and $\mathbf{S}_2$ are the basis of $\mathfrak{g}$, the metric of the space $T_p$ is *indefinite* with the signature $(1,2)$; that is, the metric matrix has one positive and two negative elements, $\mathbf{g} = \text{diag}(1, -1, -1)$.

To show it, consider transformations $\mathbf{X} \to \mathbf{a}(t) \mathbf{X} \mathbf{a}^{-1}(t)$, where $\mathbf{a}(t)$ is a one-parameter subgroup of diffeomorphisms on $G$, and $\mathbf{X}$ is an element of the tangent space $\mathbf{T}_e$ at the identity element $\mathbf{e}$ of the group $\mathbf{G}$. If $d\mathbf{a}(t)/dt|_{t=0} = \mathbf{Y}$, $\mathbf{a}(0) = 1$, differentiation on $t\mathbf{X} \to \mathbf{a}(t) \mathbf{X} \mathbf{a}^{-1}(t)$ will give $\mathbf{X} \to \mathbf{YX} - \mathbf{XY}$, $\mathbf{X} \to \text{ad}_\mathbf{X}(\mathbf{Y})$. Due to $\det(\mathbf{X}) = \det(\mathbf{a}(t) \mathbf{X} \mathbf{a}^{-1}(t))$, endomorphisms on $\mathfrak{g}$ associated with $\mathbf{X}$, $\text{ad}_\mathbf{X}$, will preserve quadratic form $|\mathbf{x}|^2 = \det(\mathbf{X})$, thus metric on $\mathfrak{g}$. Determinant $\det: \mathbf{M} \to \mathbb{R}$ is a quadratic function on matrices $\mathbf{M}$ associated with their intrinsic properties and it is a metric invariant of adjoint transformations $\mathbf{X} \to \text{ad}_\mathbf{X}$, because $\det(\mathbf{X}) = \det(\mathbf{a} \mathbf{X} \mathbf{a}^{-1})$. For arbitrary $\mathbf{S} \in \mathfrak{g}$, $\mathbf{S} = a\mathbf{S}_0 + b\mathbf{S}_1 + c\mathbf{S}_2$ over $\mathbb{R}$, $\det(\mathbf{S}) = -a^2 - b^2 + c^2$. This metric is pseudoeuclidean with signature $(1, 2)$ or $(+, -, -)$ [14, 15, 19]. The norm (or length) of the vectors of $\mathfrak{g}$ is defined as $|\mathbf{x}| \sqrt{} = |\det(\mathbf{X})|$.

The Killing form $\mathbf{k}(\mathbf{X}, \mathbf{Y})$ is another means for determining the metric on $\mathfrak{g}$. It is defined as $\mathbf{k}(\mathbf{X}, \mathbf{Y}) = \text{Trace}(\mathbf{adXadY})$. For $\{\mathbf{S}_0, \mathbf{S}_1, \mathbf{S}_2\}$, $\mathbf{k}(\mathbf{S}_i, \mathbf{S}_j) = \text{Tr}(\mathbf{adS}_i \mathbf{adS}_j)$ gives indefinite metric of the same index 1, if the signs of elements change to opposite. Its matrix form is $\mathbf{k} = 8 \text{ diag}(-1, 1, 1)$. Metrics $(+, -, -)$ and $(-, +, +)$ describe similar geometrical objects, because of the same *index* of these forms. The index is the odds of numbers of positive and negative signs.

Every algebra element $\mathbf{S}$, a matrix of linear transformations on the carrier space $\mathbf{L}$, $\mathbf{S}: \mathbf{L} \to \mathbf{L}$, is related to a vector field on $\mathbf{L}$. $\mathbf{S}$ can be transformed to the three of four possible canonical Jordan matrices. Two of them have the same view as $\mathbf{S}_0$, and $\mathbf{S}_1$ differ only by integers. There are two basic types of phase curves obtained from the matrices $\mathbf{S}_i$—*the center* (for $\mathbf{S}_0$) and *the saddle* (for $\mathbf{S}_1$, $\mathbf{S}_2$). Recall, these matrices determine functional relationships between variables shown as phase curves. The curves characteristics are related to the two types of quadratic forms: $x^2 + y^2 = r^2$ and $x^2 - y^2 = -r^2$. They are the constants of the motions along the curves hence isometries preserving some metric properties on $L$ expressed as quadratic forms—the radii of the circles in the first case, and the constant products ($\mathbf{xy} = \text{const}$) of the values of variables related to points on hyperbolas, in the second case. Because $\det(\mathbf{S}_i) = \pm 1$ the describing processes are volume preserving. $\mathbf{S}_0$ induces transformations of points on one-dimensional sphere; hence, is metric preserving. $\mathbf{S}_0$ is related to the positive definite (Euclidean) metric with signature $(0, 2)$. $\mathbf{S}_1$ ($\mathbf{S}_2$) are related to the transformations on hyperbolas and determine indefinite metric with signature $(1, 1)$. They are also metric preserving. Thus, depending on the Jordan forms to which $\mathbf{S}$ can be transformed, $\mathbf{L}_{\mathbf{s}_i}$, considered to be a two-dimensional, becomes equipped with positive definite or indefinite metrics. There is one more class related to the nilpotent matrices, which is the result of similarity transformations of $\mathbf{S}$. Because nilpotent matrices are singular, the constants of motions are stationary points on the plane, which all are the points of axis $x$. In some sense, the space of nilpotent matrices connects the first two cases.

Suppose $L$ to be a 2-dimensional linear space over $\mathbb{R}$ of biologic variables relative to the standard basis $\{\mathbf{e}_1, \mathbf{e}_2\}$. We described two types of metric on $L$ depending on Jordan forms obtained through the similarity transformations of $\mathbf{S}$. These metrics are indefinite $\mathbf{g} = \begin{pmatrix} 1 & \\ & -1 \end{pmatrix}$, related to the diagonal form $\mathbf{S}'_1 = \begin{pmatrix} m & \\ & -m \end{pmatrix}$, and Euclidean $\mathbf{g} = \begin{pmatrix} 1 & \\ & 1 \end{pmatrix}$, related to $\mathbf{S}'_0 = \begin{pmatrix} & p \\ -p & \end{pmatrix}$. Depending on $\mathbf{g}$, $L$ resembles hyperbolas or spheres. $L$ is also a two-dimensional surface, points of which are realized as conditions of two element system. These conditions are related to the points on a phase curve, which in turn, is determined by a matrix of transformations $\exp(t\mathbf{A})$ on $G$. This matrix is an element of a three-dimensional manifold $\text{SL}(2, \mathbb{R})$, $\exp(t\mathbf{A}) \in \text{SL}(2, \mathbb{R})$. At the identity element of the group $\mathbf{e} = \exp(t\mathbf{A})$, $t = 0$, there is a tangent space $T_e$, which is the Lie algebra $\text{sl}(2, \mathbb{R})$ of matrices of zero traces. Elements of the algebra form a three-dimensional vector space $L \approx \mathbb{R}^3$. Each matrix $\mathbf{A} \in \text{sl}(2, \mathbb{R})$, ($\mathbf{A} \in \mathbb{R}^3$) is a linear combination $\mathbf{A} = \Sigma a^i \mathbf{S}_i$ of three matrices $\{\mathbf{S}_0, \mathbf{S}_1, \mathbf{S}_2\}$, considered as a basis of the algebra. At each point in the space of variables $L$ matrix $\mathbf{A}$ creates a velocity vector, and through the exponential function $\mathbf{A}$ is related to the integral curves, lying on SL(2, ℝ).

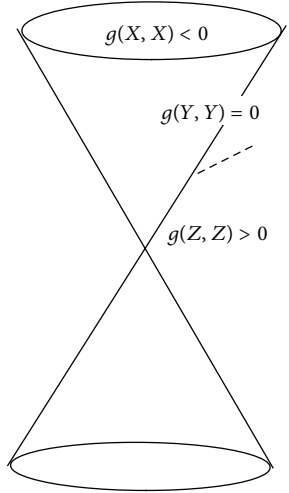

Figure 11: Indefinite metric (+, −, −) divides the space of functional elements on three subspaces: time-like, $g(X, X) < 0$, null, $g(Y, Y) = 0$, and space-like, $g(Z, Z) > 0$.

Recall, the norm of the elements of $\mathbf{L}$ is defined as $|\mathbf{S}| = \sqrt{|\det(\mathbf{S})|}$. Let the scalar product on $L(\text{SL}(2, \mathbb{R}))$ be $\mathbf{g}(\mathbf{A}, \mathbf{B}) = 4\,\text{Tr}(\mathbf{AB})$ corresponding to the Killing form. Henceforth, $L$ is a metric space. For the basis elements of the algebra (tangent space $T_e$) we have $\mathbf{g}(\mathbf{S}_i, \mathbf{S}_j) = \text{diag}(-1, 1, 1)$, where $\mathbf{g}(\mathbf{S}_i, \mathbf{S}_j) = 0$ if $i \ne j$, $\mathbf{g}(\mathbf{S}_0, \mathbf{S}_0) = -1$, $\mathbf{g}(\mathbf{S}_1, \mathbf{S}_1) = 1$, $\mathbf{g}(\mathbf{S}_2, \mathbf{S}_2) = 1$. Computations show $\mathbf{g}(\mathbf{S}_i, \mathbf{S}_j) = 4\delta_{ij}$, ($\delta_{ij}$ is a Kronecker's delta), so that $\mathbf{S}_i$ is orthogonal to $\mathbf{S}_j$ if $i \ne j$. Geometrical surface related to this metric is a hyperboloid. Indefinite metric of sl(2, ℝ) makes the space of regulatory elements inhomogeneous.

The value of the matrix determinant, $\det(\mathbf{X})$, $\mathbf{X} \in \mathfrak{g}$, is associated with the value of the scalar product $\mathbf{g}(\mathbf{X}, \mathbf{X})$. It can be positive, $\mathbf{g}(\mathbf{X}, \mathbf{X}) > 0$, negative $\mathbf{g}(\mathbf{X}, \mathbf{X}) < 0$, or zero $\mathbf{g}(\mathbf{X}, \mathbf{X}) = 0$.

According to terminology of the special relativity the vectors related to negative space-time interval or negative scalar product, $\mathbf{g}(\mathbf{X}, \mathbf{X}) < 0$, are *time-like vectors* $\mathbf{X}$, positive scalar product, $\mathbf{g}(\mathbf{Y}, \mathbf{Y}) > 0$; *space-like vectors* $\mathbf{Y}$ and zero square lengths, $\mathbf{g}(\mathbf{Z}, \mathbf{Z}) = 0$; *light-like* or *null vectors* $\mathbf{Z}$. Hence, there are three types of subspaces determined by (−1, 1, 1) metric. The subspace to which vector belongs called its *causal character* (Figure 11) [14, 16].

While indefinite metric gives all three possibilities, positive definite metric is always related to the space-like vectors, because the sum of the squared lengths is never negative. Given a span $\langle \mathbf{S}_0, \mathbf{S}_1, \mathbf{S}_2 \rangle = \mathbf{L}$, we have $\mathbf{g}(\mathbf{S}_i, \mathbf{S}_j, \mathbf{S}_k) = \text{diag}(-1, 1, 1)$. Restrictions of $\mathbf{g}(\mathbf{S}_i, \mathbf{S}_j, \mathbf{S}_k) = \text{diag}(-1, 1, 1)$ for two-dimensional spaces will give indefinite metric with signature (1, 1) and Euclidean metric (0, 2), depending on the pairs of the basis elements for two-dimensional subspaces. Let $\langle \mathbf{S}_i, \mathbf{S}_j \rangle$ be subspaces spanned by $\mathbf{S}_i$ and $\mathbf{S}_j$.

Then $\mathbf{g} \mid \langle \mathbf{S}_0, \mathbf{S}_1 \rangle = 2 \begin{pmatrix} -1 & \\ & 1 \end{pmatrix}$ and $\mathbf{g} \mid \langle \mathbf{S}_0, \mathbf{S}_2 \rangle = 2 \begin{pmatrix} -1 & \\ & 1 \end{pmatrix}$ are indefinite metrics with signature (1, 1). Vectors from the subspaces $\langle \mathbf{S}_0, \mathbf{S}_1 \rangle$ and $\langle \mathbf{S}_0, \mathbf{S}_2 \rangle$ are *time-like* vectors. Each of these subspaces contains two *null* vectors: $\mathbf{u} \pm \mathbf{v}$ for $\langle \mathbf{S}_0, \mathbf{S}_1 \rangle$, $\mathbf{g}(\mathbf{u} \pm \mathbf{v}, \mathbf{u} \pm \mathbf{v}) = 0$, and $\mathbf{u}' \pm \mathbf{v}'$ for $\langle \mathbf{S}_0, \mathbf{S}_2 \rangle$, $\mathbf{g}(\mathbf{u}' \pm \mathbf{v}', \mathbf{u}' \pm \mathbf{v}') = 0$, where $\{\mathbf{u}, \mathbf{v}\}$, $\{\mathbf{u}', \mathbf{v}'\}$ are basis vectors for $\langle \mathbf{S}_0, \mathbf{S}_1 \rangle$ and $\langle \mathbf{S}_0, \mathbf{S}_2 \rangle$, respectively.

Consider that $\mathbf{g} \mid \langle \mathbf{S}_1, \mathbf{S}_2 \rangle = 2 \begin{pmatrix} 1 & \\ & 1 \end{pmatrix}$ is a positive definite (Euclidean) and vectors from $\langle \mathbf{S}_1, \mathbf{S}_2 \rangle$ are *space-like*.

Consider $\mathbf{S}_0, \mathbf{S}_1, \mathbf{S}_2$ as representatives of subspaces $\mathscr{L}_{S_0} = \mathbb{R}\mathbf{S}_0$, $\mathscr{L}_{S_1} = \mathbb{R}\mathbf{S}_1$, and $\mathscr{L}_{S_2} = \mathbb{R}\mathbf{S}_2$. We have $\langle \mathbf{S}_i, \mathbf{S}_j \rangle = \mathscr{L}_{S_i} + \mathscr{L}_{S_j}$ and in the new notations correspondence between the metrics and the subspaces that are $\mathbf{g} = \text{diag}(-1, 1)$ is metric of time-like subspaces $\mathscr{L}^-_{S_1} = \mathscr{L}_{S_0} + \mathscr{L}_{S_1}$ and $\mathscr{L}^-_{S_2} = \mathscr{L}_{S_0} + \mathscr{L}_{S_2}$. Time-like subspaces $\mathscr{L}^-_{S_1}$ and $\mathscr{L}^-_{S_2}$ will be denoted $\mathscr{L}^-$. Time-like subspaces contain null vectors. The light-like subspace is denoted $\mathscr{L}^0$. Consider that $\mathbf{g} = \text{diag}(1, 1)$ is a metric of a space-like subspace $\mathscr{L}^+ = \mathscr{L}_{S_1} + \mathscr{L}_{S_2}$.

According to the three previously described scenarios, vector $\mathbf{S} = a\mathbf{S}_0 + b\mathbf{S}_1 + c\mathbf{S}_2$ over ℝ, being in general position, can be transformed to one of the following forms: $\mathbf{S}'_1$, $\mathbf{S}'_0$, or $\mathbf{S}'_{\text{nil}}$. This transformation rotates the tangent space $\mathscr{L}$ to have a correspondence between $\mathbf{S}'_i$ and one of the three subspaces $\mathscr{L}'_{S_0}$, $\mathscr{L}'_{S_1}$, and $\mathscr{L}'_{S_{\text{nil}}}$ (Figure 10).

Vectors from the subspace $\mathscr{L}'_{S_1}$, $\mathbf{S}'_1 = \begin{pmatrix} m & \\ & -m \end{pmatrix}$, have negative squared lengths, because of the condition $a^2 + b^2 - c^2 > 0$ for the diagonal form $\mathbf{S}'_1$, equivalent to $\det(\mathbf{S}) = \det(\mathbf{S}'_1) < 0$. $\mathscr{L}'_{S_0}$ is spanned by vectors with positive squared lengths due to $a^2 + b^2 - c^2 < 0$. These vectors have the form $\mathbf{S}'_0 = \begin{pmatrix} & p \\ -p & \end{pmatrix}$. Vectors from $\mathscr{L}'_{S_{\text{nil}}}$ are nilpotent matrices $\mathbf{S}'_{\text{nil}} = \begin{pmatrix} 0 & 1 \\ & \end{pmatrix}$ that have zero square lengths.

Direct computations of scalar products using normalized $\mathbf{S}'_0$, $\mathbf{S}'_1$, and $\mathbf{S}'_{\text{nil}}$ vectors as representatives of $\mathscr{L}'_{S_0}$, $\mathscr{L}'_{S_1}$, and $\mathscr{L}'_{S_{\text{nil}}}$ show that $\mathbf{g}'(\mathbf{S}'_0, \mathbf{S}'_1, \mathbf{S}'_{\text{nil}}) = 2\begin{pmatrix} -1 & & -1/2 \\ & 1 & \\ -1/2 & & 0 \end{pmatrix}$. ($\mathbf{g}'(\mathbf{S}'_0, \mathbf{S}'_1, \mathbf{S}'_{\text{nil}}) = 2((-1, 0, -1/2), (0, 1, 0), (-1/2, 0, 0))$ (here and further in the text in the parentheses are the matrix columns)). Consider that $\mathbf{g}' \mid \langle \mathbf{S}'_0, \mathbf{S}'_1 \rangle = 2 \begin{pmatrix} -1 & \\ & 1 \end{pmatrix}$ is indefinite; $\langle \mathbf{S}'_0, \mathbf{S}'_1 \rangle$—is a space of time-like vectors. Consider that $\mathbf{g} \mid \langle \mathbf{S}'_0, \mathbf{S}'_{\text{nil}} \rangle = 2 \begin{pmatrix} -1 & -1/2 \\ -1/2 & 0 \end{pmatrix}$ is not a degenerate metric and can be transformed to the diagonal form $\mathbf{g}'' = \text{diag}(-1, 1)$. Its signature is (1, 1), so vectors of $\langle \mathbf{S}'_0, \mathbf{S}'_{\text{nil}} \rangle$ are time-like. Consider that $\mathbf{g}' \mid \langle \mathbf{S}'_1, \mathbf{S}'_{\text{nil}} \rangle = 2 \begin{pmatrix} 1 & \\ & 0 \end{pmatrix}$ is degenerate; $\langle \mathbf{S}'_1, \mathbf{S}'_{\text{nil}} \rangle$—is a space of light-like vectors. Consider that $\mathbf{g}' = 2 \begin{pmatrix} -1 & & -1/2 \\ & 1 & \\ -1/2 & & 0 \end{pmatrix}$ is nonsingular and can be transformed over $\mathbf{S}'_1$ by $\mathbf{g}'' = \mathbf{A}\mathbf{T}\mathbf{g}'\mathbf{A}$ to the diagonal form $\mathbf{g}''(\mathbf{S}''_0, \mathbf{S}''_1, \mathbf{S}''_{\text{nil}}) = \text{diag}(-1, 1, 1)$, thus leaving $\mathbf{S}'_1$ unchanged. Consider that $\mathbf{A} = \begin{pmatrix} 1 & & 1 \\ & 1 & \\ & & -2 \end{pmatrix}$, ($\mathbf{A} = ((1, 0, 0), (0, 1, 0), (1, 0, -2))$).

Consider that $g'(S_0', S_0') = g''(S_0'', S_0'') = -1$, $g'(S_1', S_1') = g''(S_1'', S_1'') = 1$, and $g'(S_{nil}', S_{nil}') = 0$, $g''(S_{nil}'', S_{nil}'') = 1$. $S_0''$ is similar to $S_0$, $S_1''$ to $S_1$, and $S_{nil}''$ to $S_2$. Hence $\langle S_1'', S_{nil}''\rangle$ is space-like. Thus, the metric $g$ is congruent to $g''$, and $g \to g''$ is an isometry.

Analogous to the first, initial, partition, we have $\mathscr{L}_{S_1}''^{-} = \mathscr{L}_{S_0}'' + \mathscr{L}_{S_1}''$, $\mathscr{L}_{S_{nil}}''^{-} = \mathscr{L}_{S_0}'' + \mathscr{L}_{S_{nil}}''$, and $\mathscr{L}''^{+} = \mathscr{L}_{S_1}'' + \mathscr{L}_{S_{nil}}''$, where $\mathscr{L}''^{-}$ is either $\mathscr{L}_{S_1}''^{-}$ or $\mathscr{L}_{S_{nil}}''^{-}$. $\mathscr{L}^0$ is a space of light-like vectors.

It can be shown that Lorentz vector space $\mathscr{L}$, being the sum of time-like $\mathscr{L}^{-}$ ($\mathscr{L}''^{-}$), light-like $\mathscr{L}^0$ ($\mathscr{L}''^0$), and space-like $\mathscr{L}^{+}$ ($\mathscr{L}''^{+}$) subspaces, can be presented as a direct sum of just space-like and time-like subspaces $\mathscr{L} = \mathscr{L}^{-} \oplus \mathscr{L}^{+} = \mathscr{L}''^{-} \oplus \mathscr{L}''^{+}$ [14].

In biologic terms, indefinite metric with the signature $(1, 1)$ and positive definite metric $(0, 2)$ are the most consistent (stable) structural components of relationships among basis regulatory elements and the most distinguishable feature of the integrated regulatory structure of the system. $S_0$ and $S_1$ are the functional *invariants* under similarity transformations. It should be noted that metric matrices and matrices of transformations are different objects.

Indefinite metric $(-1, 1, 1)$ of the space of regulatory elements is a consequence of the structural properties of the basis elements $S_0$, $S_1$, and $S_2$ of $sl(2, \mathbb{R})$. We described indefinite metric $(-1, 1)$ also on the space $L$ of biologic variables. It was shown that $L$ can be furnished by Euclidean metric $(1, 1)$ as well.

The carrier space inherits the metric structure of regulatory elements. Each type of the basis regulatory element is fixed by some anatomical (or molecular) pairs of variables, which groups elements of the whole anatomical and molecular structure into the families of negative feedback, positive feedback, and reciprocal links. Different basis regulatory elements comprise different families; that is, a pair of anatomical structures linked by feedback cannot be regulated by reciprocal relations, and vice versa. More specifically, for example, TSH (thyroid stimulating hormone) and thyroxine (hormone of the thyroid gland) are related through the negative feedback loops, not through the reciprocal links. Therefore, it is logical to assume that there are families of 2-space $L$, isomorphic as linear spaces $\mathbb{R}^2$, but are different as biologic and geometrical objects. In fact, we have a direct sum of two-dimensional spaces representatives of each family: $L_S = L_{S_0} \oplus L_{S_1} \oplus L_{S_2}$, $L_S \cong \mathbb{R}^6$. There are convincing examples for two-dimensional spaces of variables, related by negative feedback and reciprocal links. It is proposed that positive feedback has also representation in "its own carrier space."

The matrices $S_i$ have dual features. As elements of a Lie algebra $sl(2, \mathbb{R})$, $S_i$ considered as a "one-dimensional" basis of $\mathbb{R}^3$. If $\{S_i\}$ are the elements of four-dimensional space of transformations on $\mathbb{R}^2$ whose matrix structure is related to the type of operator $S_i$, then each $S_i$ is a restriction on $\mathbb{R}^2$ and the direct sum of $S_i$, $\mathscr{S} = \oplus S_i$ will correspond to $\mathbb{R}^6$ space of $L_S$. This correspondence takes place because of induced $L_S$ structure of $\mathscr{S}$. In this case each $S_i$, being a linear combination of the unit matrix elements $\{e_1, e_2, e_3, e_4\}$ (a standard matrix basis of $\mathbb{R}^4$), belongs to $\mathbb{R}^2$. Two bases of $\mathbb{R}^4$, $\{1, S_0, S_1, S_2\}$, where $1 = \text{diag}(1, 1)$ and $\{e_1, e_2, e_3, e_4\}$, are isomorphic, and we have correspondences $S_0 = e_2 - e_3$, $S_1 = e_1 - e_4$, and $S_2 = e_2 + e_3$. Direct sum of the operators representing negative feedback and reciprocal links will certainly make sense, because $S_0$ and $S_1$ as well as their subspaces are disjunctive $\mathscr{L}_{S_0} \cap \mathscr{L}_{S_1} = 0$. Although it was proposed that positive feedback has its own representation in the carrier space $L_{S_2}$, $S_0 \cap S_2 = e_2$ (nilpotent operator $S_{nil}'$) and subspace $L_{S_2}$ (related to positive feedback) is transversal to $L_{S_0}$, but not disjunctive, $L_{S_0} \cap L_{S_2} = L_{S_{nil}}$. On the other hand, nilpotent operators $S_{nil} \neq 0$, and subspaces of light-like vectors related to $S_2$ are the physiologic background of distinguishing three orthogonal, disjunctive carrier subspaces.

The sum of the subspaces of transformations $L_{S_i}$ corresponding to the three basis matrices presents the same 3-space as a direct sum of subspaces representing negative feedback and reciprocal links: $\mathscr{L}_S = \mathscr{L}_{S_0} + \mathscr{L}_{S_1} + \mathscr{L}_{S_2}$, $\mathscr{L}_S = \mathscr{L}_{S_0} \oplus \mathscr{L}_{S_1}$, and $\mathscr{L}_{S_0} \cap \mathscr{L}_{S_2} \neq 0$. It means that $S_0$ and $S_1$ play the role of "pure" regulatory states that span $\mathscr{L}_S$: the space containing $S_1$ vectors are orthogonal to the space of $S_0$ and complements it to $\mathscr{L}_S$. Thus, the whole space $\mathscr{L}_S$ can be entirely decomposed on two subspaces, having no common regulatory components in the sense that $S_0$ and $S_1$ are disjunctive as 4-vector. In other words, regulatory actions can be a superposition of negative feedback and reciprocal links. Thus the metric of subspaces of biologic variables $L_{S_i} = \mathbb{R}^2$, related to $\mathbb{R}^3$ space of transformations $L$, can be presented as having just indefinite or positive definite metric structures. Since $S_2$ can be continually transformed to $S_1$, it seems that its functional role is to alleviate relationships between negative feedback and reciprocal links. Due to the same sign of determinants, $S_1$ and $S_2$ can be transformed to each other by continuous transformations. Unlike $S_1 \leftrightarrow S_2$, $S_1$ and $S_0$ cannot be connected by a continuous pathway. In this representation $2 \times 2$ matrices, representing $S_0$, $S_1$, $S_{nil}$, are embedded in $\mathbb{R}^{6 \times 6}$ space of transformation matrices of $\mathbb{R}^6$ space of variables as diagonal blocks. Each $2 \times 2$ diagonal block is a separable component of an integrated functional unit. Matrices of the tangent space $sl(2, \mathbb{R})$ being in general position relative to the basis $\{e_1, e_2, e_3, e_4, e_5, e_6\}$ of $L_S \cong \mathbb{R}^6$ can be transformed to the Jordan forms, so that in a new basis $\{e_1', e_2', e_3', e_4', e_5', e_6'\}$ to match one of the diagonal blocks, where blocks are $S_0$, $S_1$, and $S_{nil}$. Related to the new pair of variables $(e_i', e_j')$ the integrated subunit $M_{S_i}$ will have a block-diagonal view, where the blocks are $S_1$ or $S_0$ or $S_{nil}$ and other diagonal elements are 0s. Consider that $M_{S_0} = \text{diag}(S_1, 0, 0, 0, 0)$, $M_{S_1} = \text{diag}(0, 0, S_0, 0, 0)$, and $M_{S_{nil}} = \text{diag}(0, 0, 0, 0, 0, 0)$, whose $a_{56}$ entry, $M_{S_{nil}}(a_{56}) = 1$.

One of the most important functional determinants of the system is *stability* of its components and the whole structure. In the context of the presented materials stability is discussed in two ways, with respect to the properties of regulatory elements to be a mathematical group, and in the connection with specific features of mathematical operators and matrices themselves. The group structure of regulatory elements explains why current conditions of the system fluctuate around equilibrium states. Besides being a group, these operators have additional characteristics. For instance, an apoptosis, which is the maintenance of the equilibrium between cell proliferation and elimination, can be viewed as a distinct function fixed in reciprocal links between two opposite processes. The property of an apoptosis to prevent tissue overgrowth can formally be derived from $\mathbf{S}_1$ matrix having $\det(\mathbf{S}_1) = |1|$. For example, uncontrollable growth, known as one of the peculiar characteristics of malignancy, can be related to the indefinite metrical structure of $\mathbf{S}_1$ with the only but crucial difference; there should be $\text{Tr}(\mathbf{S}'_1) > 0$. The matrix form could be $\mathbf{S}'_1 = \begin{pmatrix} +n & \\ & -1 \end{pmatrix}$, $n > 1$. $\mathbf{S}_0$ represents negative feedback, which is similar to the reciprocal links in keeping variables within certain functional limits. This is also due to $\det(\mathbf{S}_0) = 1$. $\mathbf{S}_2$ needs a confined area, because the variables have the tendency to increase their values simultaneously and almost linearly when they come up to the asymptote. Because the phase curves of $\mathbf{S}_2$ are hyperbolas the metric properties of the phase space are preserved, if the asymptotes serve as the coordinates.

The phase curves related to $\mathbf{S}_0$, $\mathbf{S}_1$, and $\mathbf{S}_2$ are the examples of isometries preserving inner characteristics of subsystems to make regulatory process autonomous. Thus, stability of biologic systems can be described in terms of *isometries* of linear transformations. An isometry is a mapping preserving some geometrical properties of transforming objects given by the norm of vectors and operators or by the scalar product. In other words, isometries give additional characteristics to the regulatory elements depending on the type of geometrical surfaces of the carrier space and the space of transformations. For Euclidean spaces isometries are obtained by a group of orthogonal transformations, preserving, for instance, the radiuses of rotations. Isometries of the spaces with indefinite metric are related to the group of Lorentz transformations.

An isometry can be defined as a transformation $\mathbf{A}$ leaving metric tensor $\mathbf{g}$ unchanged, so that to satisfy $\mathbf{g} = \mathbf{A}^T\mathbf{g}\mathbf{A}$. For two-dimensional spaces equipped with indefinite metric, $\mathbf{g} = \begin{pmatrix} 1 & \\ & -1 \end{pmatrix}$, $\mathbf{A}: \mathbb{R}^2_1 \to \mathbb{R}^2_1$, the matrix of isometry, or Lorentz transformations is found from the above equation. It has the view $\mathbf{A} = \begin{pmatrix} \pm\text{ch}\psi & \pm\text{sh}\psi \\ \pm\text{sh}\psi & \pm\text{ch}\psi \end{pmatrix}$. This matrix represents the group of hyperbolic rotations $\mathbf{A}$ over the angle $\psi$. It transforms points lying on hyperbola to the points of hyperbola, thus leaving them on the same geometrical surface. Hyperbolas are pseudo spheres $\mathbf{S}^1_1$ (dimension = 1) with real ($a$) or imaginary ($ia$) radii satisfying $-x^2 + y^2 = a^2$ or $-x^2 + y^2 = -a^2$. For $\mathbb{R}^3_1$, correspondent surfaces are one or two chamber hyperboloids $-x^2 + y^2 + z^2 = a^2$ or $-x^2 + y^2 + z^2 = -a^2$. In other words, hyperbolic rotations preserve indefinite metric of the above quadratic forms given as the formulas of hyperbolas or hyperboloids.

Consider that $\mathbf{g}(\mathbf{S}_0, \mathbf{S}_1) = \mathbf{A}^T\mathbf{g}(\mathbf{S}_0, \mathbf{S}_1)\mathbf{A}$ gives a matrix of isometry $\mathbf{A}$, because $\mathbf{g}(\mathbf{S}_0, \mathbf{S}_1) = \begin{pmatrix} 1 & \\ & -1 \end{pmatrix}$ is preserved by Lorentz transformations $\mathbf{A}$. It is known that $\mathbf{A}$ is not a connected group and, depending on the combinations of signs of its elements, consists of four components $\mathbf{A} = \mathbf{A}^1 \cup \mathbf{A}^2 \cup \mathbf{A}^3 \cup \mathbf{A}^4$, represented by simpler forms $\mathbf{T} = \begin{pmatrix} 1 & \\ & 1 \end{pmatrix}$, $\mathbf{N} = \begin{pmatrix} -1 & \\ & -1 \end{pmatrix}$, $\mathbf{P} = \begin{pmatrix} 1 & \\ & -1 \end{pmatrix}$, and $\mathbf{NP} = \begin{pmatrix} -1 & \\ & 1 \end{pmatrix}$, respectively. It should be noted that $\mathbf{P}$ and $\mathbf{NP}$ matrices have the same view as $\mathbf{S}_1$ ($-\mathbf{S}_1$), considered previously as matrices of transformations on $\mathbf{L}$. Thus the main feature of $\mathbf{S}_1 = \begin{pmatrix} 1 & \\ & -1 \end{pmatrix}$ is reciprocal transformations of subsystems conditions, preserving metric properties of the whole space.

Isometries related to $\mathbf{S}_0$, the matrix of a negative feedback, are a group of orthogonal transformations, satisfying $\mathbf{g} = \mathbf{O}^T\mathbf{g}\mathbf{O}$; $\mathbf{g} = \begin{pmatrix} 1 & \\ & 1 \end{pmatrix}$, $\mathbf{O}: \mathbb{R}^2 \to \mathbb{R}^2$. The matrix of isometry $\mathbf{O}$ satisfying the above equation is $\mathbf{O} = \begin{pmatrix} \cos\phi & -\sin\phi \\ \sin\phi & \cos\phi \end{pmatrix}$. The matrix $\mathbf{O}$ ($\det(\mathbf{O}) = 1$) belongs to the group of proper rotations of the plane through the angle $\phi$ preserving Euclidean metric $\mathbf{g}$ that gives concentric circles on the phase space of two variables. Consider that $\mathbf{g}(\mathbf{S}_1, \mathbf{S}_2) = \mathbf{O}^T\mathbf{g}(\mathbf{S}_1, \mathbf{S}_2)\mathbf{O}$ gives an orthogonal isometry matrix $\mathbf{O}$ with the property $\det(\mathbf{O}) = 1$. $\mathbf{O}$ determines rotations preserving Euclidean metric.

Consider $\text{SL}(2, \mathbb{R})$ to be a space of transformations on $\mathbf{L}$ of biologic variables, as a metric space. Its metric tensor, defined at each point of $\text{SL}(2, \mathbb{R})$, has a signature (1, 2) inherited from the indefinite metric of the tangent spaces to these points. Tangent spaces are the algebra $\text{sl}(2, \mathbb{R})$ spanned by the basis $\{\mathbf{S}_0, \mathbf{S}_1, \mathbf{S}_2\}$. Because for each $\mathbf{a} \in \text{SL}(2, \mathbb{R})$ $\det(\mathbf{a}) = 1$, and inner endomorphisms of $\text{SL}(2, \mathbb{R})$, $I_\mathbf{a} = \mathbf{f} = \mathbf{asa}^{-1}$, give elements of the group $\mathbf{f} \in \text{SL}(2, \mathbb{R})$, $\det(\mathbf{f}) = 1$, $I_\mathbf{a}$ preserve determinant function $\det(\mathbf{f}) = \det(\mathbf{a}^{-1}\mathbf{sa}) = 1$, thus can be considered as the *volume preserving isometries*.

## 8. Regulatory Functions of a Biologic System as Inhomogeneous Space of Metric Matrices with Preserved Isometry

Regulatory structure of biologic system can also be considered in a way, when regulatory elements are represented by metric matrices. For these purposes $\mathscr{L}$ is realized as a three-dimensional space of metric matrices. Matrices of transformations $\mathbf{S}_0$, $\mathbf{S}_1$, and $\mathbf{S}_2$ become Gramm's matrices, denoted $\mathbf{S}_0$, $\mathbf{S}_1$, and $\mathbf{S}_2$ of bilinear forms. Each $\mathbf{S}_i$ has its own metric structure: $\mathbf{S}_1$, $\mathbf{S}_2$ are indefinite metric matrices; $\mathbf{S}_0$ is symplectic matrix. $\mathbf{S}_1$, $\mathbf{S}_2$ quadratic forms on $\mathscr{L}$ have the same features as in cases where $\mathbf{S}_1$, $\mathbf{S}_2$ are transformation matrices, whereas $\mathbf{S}_0$, the symplectic metric, can be considered as a generalization of rotations in Euclidean spaces. Symplectic metric does not "separate" variables, so regulatory function, for example, in $\mathbb{R}^2$, can be applied only to pairs of variables. Scalar product of one-dimensional objects related to symplectic metric is zero; $(\mathbf{v}, \mathbf{S}_0\mathbf{v}) = 0$. On the contrary, indefinite metric distinguishes the class of the one-dimensional objects, so that the division of a carrier space can be reduced to the invariant "autonomous" subspaces of lesser dimensions.

Metric $\mathbf{X} \in \mathscr{L}$ can be determined equivalently through the determinant function $\det(\mathbf{X})$ or the Killing form. The basis $\{\mathbf{S}_i\}$ endows $\mathscr{L}$ with the indefinite metric $\mathbf{g} = \text{diag}(-1, 1, 1)$, so that $\mathscr{L}$ is a three-dimensional Lorentz vector space $(-, +, +)$.

Metric matrices $\mathbf{S} = \sum a^i \mathbf{S}_i$ can be transformed either to $\begin{pmatrix} k & \\ & -k \end{pmatrix}$ or $\begin{pmatrix} & m \\ -m & \end{pmatrix}$ forms, so that resulting hyperbolic or symplectic spaces become projections of initial metric structure on corresponding two-dimensional carrier spaces. Depending on the basis of **L**, hyperbolic space can be presented by $\begin{pmatrix} k & \\ & -k \end{pmatrix}$ or $\begin{pmatrix} & p \\ p & \end{pmatrix}$ matrices.

In general, isometries of hyperbolic and symplectic spaces can be characterized by transformations preserving the forms of corresponding metric matrices: $\mathbf{S}_1 = \mathbf{A}^T\mathbf{S}_1\mathbf{A}$ gives matrix of isometry **A**, because $\mathbf{S}_1$ is preserved by **A**; $\mathbf{S}_0 = \mathbf{M}^T\mathbf{S}_0\mathbf{M}$ is related to the symplectic matrix **M** of isometry with the property $\det(\mathbf{M}) = 1$. It has the same view as $\mathbf{S}_0$. Transformations preserving $\mathbf{S}_2$ follow from $\mathbf{S}_2 = \mathbf{B}^T\mathbf{S}_2\mathbf{B}$, where **B** can be represented by the matrix of the same view as $\mathbf{S}_2$.

Metric properties of the carrier spaces $\mathbf{L}_{\mathbf{S}_i} = \mathbb{R}^2$ of biologic variables are inherited from metrics of corresponding regulatory structure. $\mathbf{L}_{\mathbf{S}_1}$ and $\mathbf{L}_{\mathbf{S}_2}$, related to reciprocal links and positive feedback, have indefinite metric structure, whereas $\mathbf{L}_{\mathbf{S}_0}$, related to negative feedback, has symplectic structure.

Metric properties of **S** can be summarized in the law of conservation of energy, expressed through the ordinary differential equation $d\mathbf{x}/dt = (a\mathbf{S}_0 + b\mathbf{S}_1 + c\mathbf{S}_2)\mathbf{x}$, where $a$, $b$, and $c$ are real numbers. This equation shows that change in the state **x** of biologic system in a time interval $t$ is determined by the integrated vector field **S**. This equation describes autonomous system not explicitly depending on $t$, and it is a linear approximation of "real" processes.

Each $\mathbf{S}_i$ corresponds to the vector field related to the constant first integral (or constant energy level); therefore integrated field $\mathbf{S} = a\mathbf{S}_0 + b\mathbf{S}_1 + c\mathbf{S}_2$ over $\mathbb{R}$ will determine a conservative, closed system with the constant inner energy [16, 19, 23]. Depending on the values of coefficients $a$, $b$, $c$, **S** can be transformed to $\mathbf{S}_0$, $\mathbf{S}_1$, or $\mathbf{S}_{\text{nil}}$ matrices giving three types of solutions corresponding to the new variables.

## 9. Conclusion

Initially the purpose of this paper was to validate and describe the basis functional components of the biologic system, which makes the system a closed structure under the existing regulatory mechanisms. Later on, the Lie algebra formalism led to the metric properties of the space of regulatory elements of the biologic system. It seemed logical to include two closely related results in this paper.

Presented conception of a biologic system is based on a general approach describing regulatory structure of the whole system through the quantification of its functional parts. It is proposed that the basis functional elements resembling universal parts of regulatory structure (subsystems) have become separable during phylogenetic and ontogenetic development, when a manifestation of new functional properties and appearance of new morphological structures were accompanied by three (specific for structural pairs) types of regulatory mechanisms. In two element systems the three regulatory basis elements, positive feedback, negative feedback, and reciprocal links, form an *integrative regulatory unit*. Acting simultaneously, they determine current conditions of the system.

This conception is based on the structure of the basis elements of a Lie algebra $sl(2, \mathbb{R})$ and the related group $SL(2, \mathbb{R})$, having the properties to be closed structures under compositions of their elements. Three basis elements of the algebra, $\mathbf{S}_0 = \begin{pmatrix} & 1 \\ -1 & \end{pmatrix}$, $\mathbf{S}_1 = \begin{pmatrix} 1 & \\ & -1 \end{pmatrix}$, and $\mathbf{S}_2 = \begin{pmatrix} & 1 \\ 1 & \end{pmatrix}$, represent two well-known regulatory mechanisms, negative and positive feedback, and the newly described reciprocal links. Together they form integrative vector field $\mathbf{S} = a\mathbf{S}_0 + b\mathbf{S}_1 + c\mathbf{S}_2$ over $\mathbb{R}$ on the space of biological variables. Each regulatory component $\mathbf{S}_i$ "phylogenetically" is fixed on its own 2-space of biologic variables, so that together they form a morphofunctional six-dimensional integrative unit. In practice we understand it as if the whole 6-space is reduced to the two-dimensional subspaces (subsystems) regulated either by positive feedback, negative feedback, or reciprocal links.

This provides the clue to new understanding of the fact that the regulatory components of the system should form a closed functional structure. This is the key point for keeping the system stable.

Adjoint representations of the group and algebra elements ($\mathbf{f} \rightarrow \mathbf{Ad}(\mathbf{f})$ and $\mathbf{S} \rightarrow \mathbf{ad}(\mathbf{S})$) show the universal character of regulatory elements on different functional levels—regulation of metabolic processes and autoregulation of functional elements themselves. The basis elements $\mathbf{S}_i$ of $sl(2, \mathbb{R})$ and related vector fields represent their characteristics in the ability of the corresponding subsystem to maintain their internal structure and autonomy. Thus each regulatory subsystem can be viewed as conservative and stable, capable of maintaining its inner energy, that is, to be independent of the external sources of energy. This simplification helps in understanding the fact that, despite the differences in functional properties of regulatory components, the group properties endow the integrated structure with the same features as the components. This makes biologic systems uniformly organized on different hierarchical levels.

From the structure of $\{\mathbf{S}_0, \mathbf{S}_1, \mathbf{S}_2\}$ basis of the Lie algebra it formally follows that all events regarding regulatory processes of normal biologic systems, geometrically speaking, should lie on some geometrical surface, which in our case is hyperboloid.

Primary regulatory functions $(\mathbf{S}_0, \mathbf{S}_1, \mathbf{S}_2)$ comprise three-dimensional space having indefinite metric structure with signature $(1, 2)$ or $(+, -, -)$. It resembles relativistic Minkowski space-time restricted to two spatial coordinates. Geometrically, regulatory processes are trajectories on the surface resembling hyperboloids. Scalar product, considered as a measure of functional activities of regulatory elements themselves or their interactions, divides the 3-space of regulatory elements on three subspaces containing "time-like," "null," and "space-like" vectors. Self-preservation of this metric and its components is an essential feature of biologic organization.

Indefinite metric structure provides functional and morphological flexibility to biologic systems, which could be reflected in the system's ability to be highly adapted to the constantly changing environment.

The meaning of indefinite metric for biologic objects is based on reciprocity, which is an efficient means in achieving equilibrium states. Formally, the reciprocal links directly reflect indefinite metric in the regulatory structure.

The basis elements can be transformed into each other, which can be extrapolated on the ability of real systems to combine potent regulatory mechanisms in order to obtain alternative pathways to compensate deteriorated function. Structural "deterioration" of any of the three basis components may affect the system's steadiness and the ability to maintain genetically fixed relations among regulatory components. Practically, it may cause either an inadequate stimulation or suppression of regulatory functions, leading to excessive activity or degeneration. It can be presented in the form of uncontrollable cell proliferation. Most probably, this process has a more complex underlying mechanism, related to the deterioration of all three basis regulatory components.

This concept can be generalized to $n$-dimensional spaces. Almost all the statements, discussed in mathematical terms, can be easily interpreted using physiological language. This approach has a broad spectrum of possibilities for further theoretical and practical investigations. For example, it may boost to find out how the reciprocal links are related to already known mechanisms of regulations, what structural changes in the basis regulatory mechanisms determine functional and morphological deterioration, leading, for instance, to the uncontrollable and invasive cell proliferation. The latter is a clue to understand the functional phenomenon of cancer growth as a "deviation" from the uniformly organized regulatory structure of biologic systems.

## Highlights

The mathematical model of organization and self regulation of biologic systems gives insight to the most common aspects of biologic organisation. Positive feedback, negative feedback, and reciprocal links expressed in a matrix form are identical to the basis elements of Lie algebra sl(2, $\mathbb{R}$); thus, they may represent a functional basis of biologic system. Three basis elements form closed functional structure maintaining stability of the system. The space of biologic variables is inhomogeneous and has indefinite metric structure.

## Conflict of Interests

The author declares that there is no conflict of interests regarding the publication of this paper.

## Acknowledgment

The author would like to thank Dr. Janetta Kourzenkova for discussions and useful suggestions made while preparing this material. This work was not supported by grants.

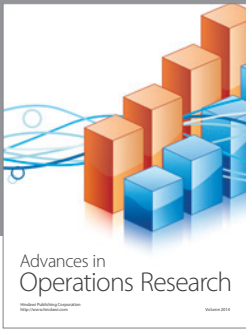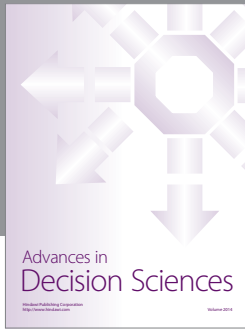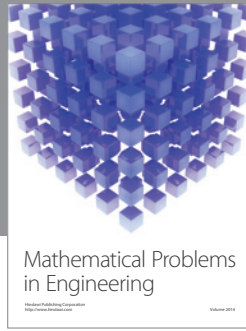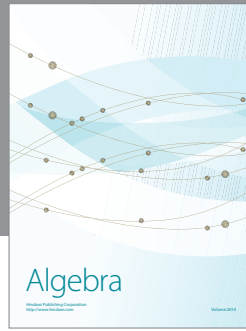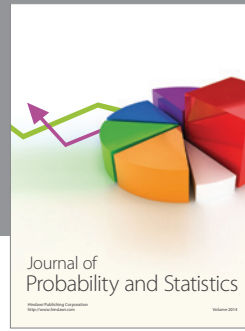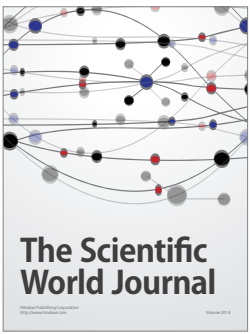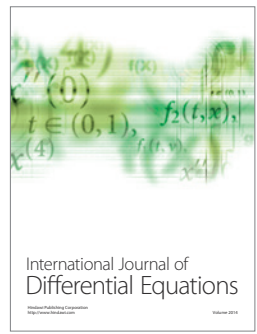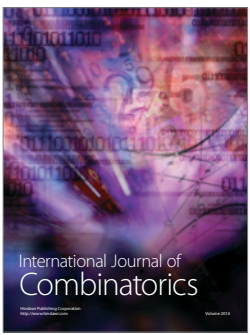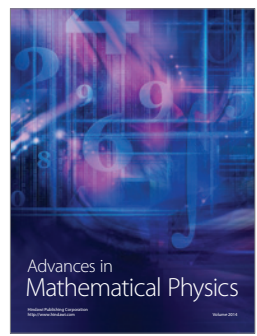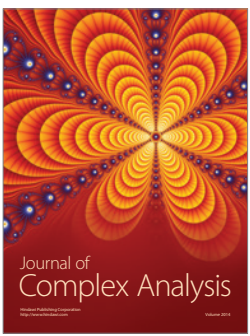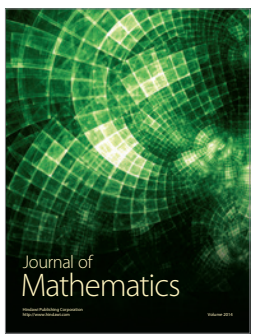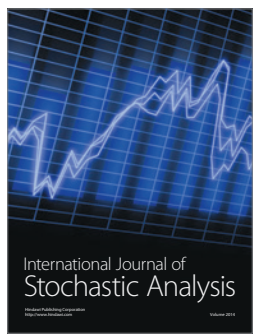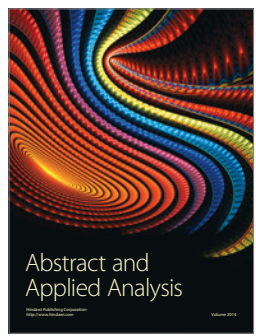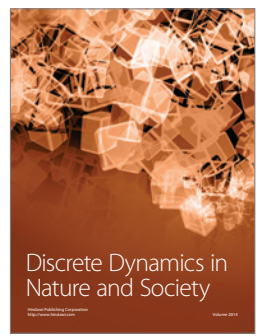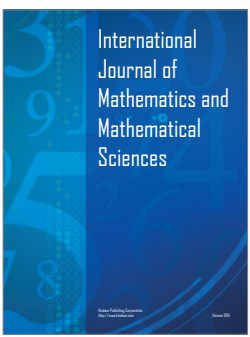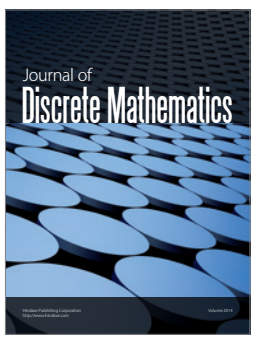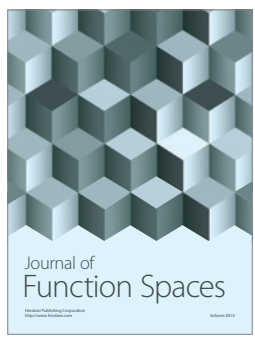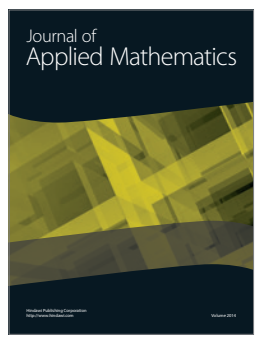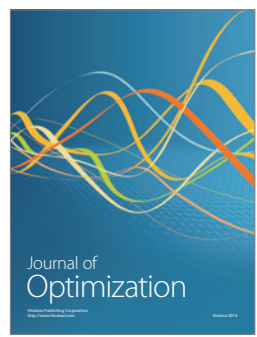

Submit your manuscripts at
http://www.hindawi.com